\documentclass{article}%
\usepackage{amssymb}
\usepackage{amsfonts}
\usepackage{amsmath}
\usepackage{graphicx}
\usepackage{caption}
\usepackage{subcaption,xcolor}
\usepackage{lipsum}
\usepackage{chngcntr}
\counterwithin{figure}{section}
\providecommand{\U}[1]{\protect\rule{.1in}{.1in}}
\newtheorem{theorem}{Theorem}[section]

\newtheorem{corollary}[theorem]{Corollary}
\newtheorem{definition}{Definition}[section]

\newtheorem{lemma}[theorem]{Lemma}
\newtheorem{proposition}[theorem]{Proposition}

\newcommand\blfootnote[1]{%
	\begingroup
	\renewcommand\thefootnote{}\footnote{#1}%
	\addtocounter{footnote}{-1}%
	\endgroup
}
\begin{document}
	
\title{Optimal dividends for a NatCat insurer \\ in the presence of a climate tipping point}
\author{Hansj\"{o}rg Albrecher\thanks{Corresponding Author. Department of Actuarial Science, Faculty of
		Business and Economics,  Swiss
		Finance Institute and Expertise Center for Climate Extremes (ECCE), University of Lausanne, CH-1015 Lausanne. Email: hansjoerg.albrecher@unil.ch}, Pablo Azcue\thanks{Departamento de Matematicas, Universidad
		Torcuato Di Tella. Av. Figueroa Alcorta 7350 (C1428BIJ) Ciudad de Buenos
		Aires, Argentina. Email: pazcue@utdt.edu and nmuler@utdt.edu}\;\footnote{Pablo Azcue contributed to the present manuscript before he unfortunately passed away in April 2024. We dedicate this work to him.} and Nora Muler$^{\dag}$}
\date{}
\maketitle

\bigskip\abstract{\begin{quote}
		\noindent 
	{We study optimal dividend strategies for an insurance company facing natural catastrophe claims, anticipating the arrival of a climate tipping point after which the claim intensity and/or the claim size distribution of the underlying risks deteriorates irreversibly. Extending earlier literature based on a shot-noise Cox process assumption for claim arrivals, we show that the non-stationary feature of such a tipping point can, in fact, be an advantage for shareholders seeking to maximize expected discounted dividends over the lifetime of the portfolio. Assuming the tipping point arrives according to an Erlang distribution, we demonstrate that the corresponding system of two-dimensional stochastic control problems admits a viscosity solution, which can be numerically approximated using a discretization of the current surplus and the claim intensity level. We also prove uniform convergence of this discrete solution to that of the original continuous problem. The results are illustrated through several numerical examples, and the sensitivity of the optimal dividend strategies to the presence of a climate tipping point is analyzed. In all these examples, it turns out that when the insurance premium is adjusted fairly at the moment of the tipping point, and all quantities are observable, the non-stationarity introduced by the tipping point can, in fact, represent an upward potential for shareholders.}
\end{quote}}	

\section{Introduction\label{Section Introduction}}
In the last years, it became amply clear that climate change leads to more frequent and more severe weather extremes, see e.g.\ Sneviratne et al.\ \cite{Seneviratne}, thus leading to an increase of natural catastrophes, cf.\ e.g.\ Lloyd and Shepherd \cite{Lloyd}. Correspondingly, the insurance of losses due to natural catastrophes (NatCat) is becoming increasingly challenging, see e.g.\ Schneider \cite{Schneider}. Traditionally, the reinsurance market was able to provide protection for respective lines of business to quite some extent (see \cite{ABT17} for a survey on reinsurance from an actuarial perspective), but due to these recent developments reinsurance premiums for several NatCat perils are rising, causing increasing concern in the insurance industry. In this context it is of interest to see whether the {insurance of NatCat lines remains} feasible and/or attractive for insurance companies. Among several different angles from which one might want to study such a question, in \cite{AAM24} we recently looked into the insurance of NatCat risks from a shareholder perspective. In particular, we studied whether the variability of claim intensities over time, which naturally appears in NatCat lines, can also have upward potential for shareholders when performance is measured in terms of the expected value of the aggregate discounted future dividends from the insurance portfolio, a criterion that goes back to de Finetti \cite{defin} and that has been studied in terms of optimal respective dividend strategies in risk theory ever since (see e.g.\ \cite{AT,avanzi} for surveys). {In \cite{AAM24}, it was recently shown that if the Poisson claim intensity in a NatCat portfolio is modeled by a (stochastic) shot-noise process and this process is observable, then the variability of the claim intensity can indeed enhance the value of the insurance portfolio -- provided the dividend strategy is suitably adapted to the evolution of the observed claim intensity over time.} For the solution of the respective two-dimensional optimal control problem, it was necessary to assume a certain degree of stationarity of this shot-noise process, {restricting} the considerations to NatCat portfolios for which neither the intensity nor the claim severity distribution {could} systematically increase over time. \\
In this paper we would like to go one step further and allow for such a systematic change (in the sense of a deterioration) in the analysis. 
{To ensure that the typical stochastic control techniques from previous studies remain applicable, we embed this extension within a Markovian framework by assuming that a 'change-point' in the conditions occurs at a stochastic time, driven by an independent renewal process with specified characteristics.} Concretely, we consider a Cox process for the claim arrivals due to a shot-noise process as in \cite{AAM24}, but at an exponentially (or Erlang) distributed time the dynamics of the process change to a more severe situation. While the developed technique will{, in principle, allow for several} such changes over time, we restrict the considerations to one such occurrence in this paper, which will simplify the presentation and numerical implementation when deriving the optimal dividend strategy. One can view such a sudden change as a discrete and mathematically tractable approximation of an increasing trend. On the other hand, such a 'change-point' also has a natural interpretation in terms of a \textit{climate tipping point}, which is a critical threshold that, once crossed, leads to an irreversible change in climatic conditions. This can happen in response to a gradual warming of air and ocean surface temperature. Over the last years, an increasing number of climate scientists {predicted such} a tipping point to occur with a large probability (see e.g.\ Lenton et al.\ \cite{LentonTipping}), and we choose to focus on the tipping point interpretation of our model in the rest of the paper. Here we assume that after the occurrence of this tipping point we will have a higher intensity of claims and/or a change in the claim severity distribution. Our goal is to see that, if one is able to actually observe the occurrence of such a tipping point and has realistic assumptions on the implications {for} the risk characteristics of a NatCat portfolio, it is possible to derive an optimal dividend strategy for an insurance portfolio exposed to this risk. We assume independent and identically distributed claims, but with a potentially different distribution before and after the tipping point. The premiums for the policyholders are  calculated according to an expected value principle (with a safety loading), which stays constant before the tipping point (so that the risk of changed claim intensities due to the arrival of catastrophes is a priori borne by the insurance company and diversified over time), but at the time of the tipping point this premium is adapted to its new level due to the systematically increased risk (and then kept constant again at that higher level). \\

The setup will allow {us} to solve the optimal dividend problem as an iteration of two-dimensional optimal control problems, where one variable is the current surplus of the portfolio and the other is the current claim intensity level. Concretely, we will adapt the results of \cite{AAM24} to first solve an auxiliary optimization problem that consists of determining the optimal dividend strategy for a compound Cox process with shot-noise intensity when the time horizon is limited to an exponential time. The solution of this problem can then be concatenated to provide the general solution for the optimal dividend problem in the presence of the tipping point. As it will turn out, also in this situation, the expected discounted dividends for the shareholders increase in case the processes are observable and the premium adaptation is done instantaneously. \\

On the conceptual level, there are some similarities {between the present paper and} dividend problems in Markov-modulated risk models, see e.g.\ Wei et al.\ \cite{WYW2}, Jiang and Pistorius \cite{Jiang} and Szölgyenyi \cite{Szölgyenyi}, but here we only switch in one direction. On the other hand, our underlying compound Cox risk model makes the analysis of the stochastic control problem much more complicated, and the exposition is necessarily relatively technical. We will also deal with a situation where the tipping point arrives at an Erlang($n$) time, which introduces memory into the arrival process of the tipping point, {but is still computationally tractable}. Technically this can be handled as $n$ consecutive switches at independent exponential times, an idea {that goes back, in mathematical finance, to} Carr \cite{Carr} and in actuarial science to Asmussen et al.\ \cite{AAU}, and which was already {used} in another context of dividend optimization in Albrecher and Thonhauser \cite{AT12}.\\

Note that the method we employ here can also be useful in other settings, where one needs to solve iterative optimal control problems and the solution of the previous problem {serves as} a boundary value of the next one. While the computational complexity might increase when one wants to apply multiple iterations of such a procedure, it {is still feasible methodologically.}  In that vein, one could for instance approximate a deterministic time horizon for a change of the model dynamics by an Erlang($n$) distribution with higher $n$ at the expense of $n$ iterated stochastic problems, and one could also introduce several (instead of one) 'deterioration checkpoints' of environmental conditions in the insurance activity. More generally, the method proposed in this paper might be useful in other contexts where for computational reasons one would like to approximate some non-stationary phenomenon by iterative applications of Markovian sections which pass on their result as a boundary condition to the next level.   \\

The rest of the paper is organized as follows. Section \ref{Model and basic results} formulates an auxiliary optimal control problem that will be used later as a building block for the general model. {This auxiliary problem setting with a pre-defined exit value at a stopping time may also be of interest in its own right for other modelling circumstances.} Some basic properties of the respective value function are derived. Section \ref{Seccion Viscosidad} then derives the Hamilton-Jacobi-Bellman (HJB) equation of this auxiliary problem and shows that the optimal value function is the smallest viscosity supersolution of the HJB equation. Section \ref{Section Asymptotic} provides an important boundary result as well as an asymptotic property of this value function. In Section \ref{Numerical Scheme} a numerical scheme is introduced that approximates the optimal value function on a finite grid in two-dimensional space, and it is shown that this approximation converges uniformly to the optimal value function. In Section \ref{Climate Tipping Point}, the {auxiliary model studied in the previous sections is then} applied to solve a dividend problem for a NatCat insurer {like} the one in \cite{AAM24}, but {now} with the presence of a climate tipping point. In Section \ref{Numerical Results} we then work out several concrete numerical illustrations that extend the ones in \cite{AAM24} by the presence of such a climate tipping point and we interpret the results. Finally, Section \ref{concl} concludes. All proofs {are in the} Appendix.

\section{An auxiliary optimization problem\label{Model and basic results}}

{Let us introduce an auxiliary optimization problem. It will be a crucial ingredient to inductively solve the problem of optimizing expected cumulative discounted dividend payouts in the presence of a climate tipping point, as described in Section 6.}\\ We assume that the uncontrolled surplus of the company follows a
compound Cox process with drift, generated by a shot-noise intensity until a
stopping time given by a random exponential time $\overline{\tau}\sim
\text{Exp}(\xi)$ independent of the compound Cox process. That is, for $t<\overline
{\tau},$%
\begin{equation}
	X_{t}=x+pt-\sum_{j=1}^{N_{t}}U_{j}\text{,}\label{UncontrolledSurplusOriginal}%
\end{equation}
where $x$ is the initial surplus of the portfolio and $p$ is the premium rate. Moreover, let us denote by $\tau_{j}$ the arrival
time of claim $j$. We assume that the counting process
\[
N_{t}=\#\{j:\tau_{j}\leq t\}
\]
is an inhomogeneous Poisson process with (stochastic) intensity $\lambda_{t}$. The
process $N_{t}$\ and the claim sequence $\{U_{i}\}_{i\ge 1}$ are independent of each other. The intensity
$\boldsymbol{\lambda}=(\lambda_{t})_{t\geq0}$ is a shot-noise
process given by
\begin{equation}
	\lambda_{t}=\lambda_{t}^{c}+\sum_{k=1}^{\widetilde{N}_{t}}Y_{k}~e^{-d(t-T_{k}%
		)},\label{Lambda t corrida}%
\end{equation}
where {$T_{k}$ is the arrival time of catastrophe $k$ which 
	generates an upward jump in the intensity process of size 
	$Y_{k}$. The positive random variables $\left(  Y_{k}\right)  _{k\geq1}$ are
	assumed to be i.i.d.\ (and independent of $\widetilde{N}_{t}$) with distribution function
	$F_{Y}$ and finite expectation. The arrival process of catastrophies 
	\[
	\widetilde{N}_{t}=\#\{k:T_{k}\leq t\}
	\]
	is a Poisson process of constant intensity $\beta$, independent of the current intensity value. After the arrival of a catastrophe, the induced intensity decays deterministically and exponentially with rate $d>0$. $\underline{\lambda}>0$ is a baseline intensity and 
	\begin{equation}
		\lambda_{t}^{c}=\underline{\lambda}+e^{-dt}\left(  \lambda-\underline{\lambda
		}\right)  \label{Lambda continua}%
	\end{equation}
	is the sum of that baseline intensity and the deterministic decay function of the initial intensity level $\lambda=\lambda_0\geq\underline{\lambda}$ (see for instance \cite{AA06} for such a model assumption in risk theory). }
Note that $\lambda_{t}%
\geq\underline{\lambda}$ for all $t\geq0$ and if the initial intensity satisfies
$\lambda>\underline{\lambda}$, then $\lambda_{t}>\underline{\lambda}$ for all
$t\geq0$. The baseline $\underline{\lambda}$ can be interpreted as the claim intensity of 'regular' non-catastrophic claims in the portfolio. {All quantities are considered in a suitable compound Cox filtered
space \textit{\ }$(\Omega^{\mathbf{\lambda}},\mathcal{F}^{\mathbf{\lambda}%
},\left(  \mathcal{F}_{t}^{\mathbf{\lambda}}\right)  _{t\geq0},\mathbb{P}%
^{\mathbf{\lambda}})$ conditional on the process $\boldsymbol{\lambda}$, for details see \cite[Sec.2]{AAM24} and also to Dassios and Jang \cite{Dassios Jang}.}

The company pays dividends to the shareholders up to $\overline{\tau}$ or the
ruin time (whichever comes first). We denote the dividend strategy as
$L=\left(  L_{t}\right)  _{t\geq0}$, where $L_{t}$ is the amount of cumulative dividends
paid up to time $t$. To extend the definition of $L_{t}$ for all
$t\geq0$, let us consider $L_{t}$ constant after the minimum between
$\overline{\tau}$ and the ruin time. The dividend payment strategy is
admissible if it is non-decreasing, c\`{a}dl\`{a}g (right-continuous with left
limits), adapted to $\left(  \mathcal{F}_{t}^{\mathbf{\lambda}%
}\right)  _{t\geq0}$, and satisfies $L_{0}\geq0$ and $L_{t}\leq X_{t}$ for
$t<\tau^{L},$ where the ruin time $\tau^{L}$ is defined as \ \
\begin{equation}
	\tau^{L}=\inf\left\{  t\geq0:X_{t}-L_{t^{-}}<0\right\}  \text{.}%
	\label{DefinicionTauRuina}%
\end{equation}
This last condition means that ruin can only happen at the arrival of
a claim and that no lump sum dividend payment can be made at the time of ruin. We
define the controlled surplus process as%
\begin{equation}
	X_{t}^{L}=X_{t}-L_{t}\text{.}\label{Controled Process}%
\end{equation}
Denote by $\Pi_{x,\lambda}$ the set of admissible dividend strategies
starting with initial surplus level $x\geq0$ and initial intensity
$\lambda\geq\underline{\lambda}.$ Let us define for a function $u:\left[
0,\infty\right)  \times\left[  \underline{\lambda},\infty\right)
\rightarrow\left[  0,\infty\right)  $, the growth condition
\begin{equation}
	u(x,\lambda)\leq K+x\text{ for all }(x,\lambda)\in(0,\infty)\times
	(\underline{\lambda},\infty)\text{ for some }K>0\text{.}%
	\label{Growth Condition}%
\end{equation}

We consider a function $g:\left[  0,\infty\right)  \times\left[
\underline{\lambda},\infty\right)  \rightarrow\left[  0,\infty\right)  $ (which
represents in the auxiliary model the exit value of the company at time
$\overline{\tau} $) and we study the problem of maximizing the expected
cumulative discounted dividend payouts. We assume the following conditions on
$g$:\\

\textbf{Assumption (A1):} $g$ is uniformly continuous in $\left[  0,\infty\right)
\times\left[  \underline{\lambda},\infty\right)  ,$ locally Lipschitz\ in
$(0,\infty)\times(\underline{\lambda},\infty)$, non-decreasing in $x$,
non-increasing in $\lambda$ ,%
\[
\lim_{\lambda\rightarrow\infty}\sup_{x}g(x,\lambda)=x,
\]
$g_{x}\geq1$ a.e. and $g(x,\lambda)=(x-\widehat{x})+g(\widehat{x},\lambda)$
for some $\widehat{x}\geq0$.\\

Assumption (A1) for the function $g$ collects in fact the properties of the optimal
value function that maximizes the expected cumulative discounted dividend
payments in the case that the uncontrolled surplus of the company follows a
compound Cox process with drift generated by a shot-noise intensity until
\ ruin without a switching time (see Albrecher et al.\ {\cite{AAM24}}) and these are the natural assumptions for the climate tipping point problem
(see Section \ref{Climate Tipping Point}).

For any initial surplus level $x\geq0$ and initial intensity $\lambda
\geq\underline{\lambda}$, we define the optimal value function as%
\begin{equation}
	V(x,\lambda)=\sup_{L\in\Pi_{x,\lambda}}J(L;x,\lambda),\label{DefinitionV}%
\end{equation}
where%

\begin{equation}
	J(L;x,\lambda)=\mathbb{E}\left(
	\int\nolimits_{0^{-}}^{\overline{\tau}\wedge\tau^{L}}
	e^{-qs}dL_{s}+e^{-q(\overline{\tau}\wedge\tau^{L})}I_{\{\overline{\tau}%
		\wedge\tau^{L}<\tau^{L}\}}g(X_{\overline{\tau}\wedge\tau^{L}}^{L}%
	,\lambda_{\overline{\tau}\wedge\tau^{L}})\right)\text{.}\label{Definicion VL}%
\end{equation}
Here $q>0$ is a constant discount factor. We assume that the premium rate $p$
is obtained using the expected value principle of the asymptotic distribution
of $\lambda_{t}$. That is%

\begin{equation}
	p=(1+\eta)\mathbb{E(}U_{1}\mathbb{)\lambda}_{\text{av}},\label{Definicion p}%
\end{equation}
where $\eta>0$ is the safety loading,%
\begin{equation}
	\Lambda_{t}=\int_{0}^{t}\lambda_{s}ds\label{Integral de Lambda}%
\end{equation}
and%
\begin{equation}
	\mathbb{\lambda}_{\text{av}}:=\lim_{t\rightarrow\infty}\mathbb{E}(\frac{\Lambda_{t}%
	}{t})=\underline{\lambda}+\dfrac{\beta~\mathbb{E}(Y_{1})}{d}.\label{LambdaAv}%
\end{equation}

The optimal value function $V$ will inherit the basic properties from $g$ as
we will show in the next proposition and in Section \ref{Section Asymptotic}.

In the remainder of this section, we prove some elementary properties of the
optimal value function. The proofs of the propositions in this section are given in
Appendix \ref{Appendix Section Model and Basic Results}.

\begin{proposition}
	\label{Basic Results V} $V$ is non-increasing in $\lambda,$ uniformly
	continuous in $\left[  0,\infty\right)  \times\left[  \underline{\lambda
	},\infty\right)$, locally Lipschitz\ in $[0,\infty)\times
	(\underline{\lambda},\infty)$, non-decreasing in $x$ and satisfies growth
	condition (\ref{Growth Condition})$.$
\end{proposition}

\noindent We now state the so-called Dynamic Programming Principle (DPP), but skip its
proof as it is similar to Lemma 3.6 of {\cite{AAM24}} and Lemma
1.2 of {\cite{AM 2014}.}

\begin{lemma}
	\label{DPP}For any initial surplus $x\geq0$ and any finite stopping time $\tau$, we
	can write
\begin{align*}
	&  V(x,\lambda) =\sup\nolimits_{L\in\Pi_{x,\lambda}}\left(  \mathbb{E}(%
	{\textstyle\int\nolimits_{0^{-}}^{\tau\wedge\overline{\tau}\wedge\tau^{L}}}
	e^{-qs}dL_{s}+I_{\{\overline{\tau}\wedge\tau\wedge\tau^{L}=\overline{\tau}%
		\}}g(X_{\overline{\tau}}^{L},\lambda_{\overline{\tau}})+I_{\{\overline{\tau
		}\wedge\tau\wedge\tau^{L}=\tau\}}e^{-q\tau}V(X_{\tau}^{L},\lambda_{\tau
	})\right)  .
\end{align*}
\end{lemma}%

\noindent In the next proposition we show the asymptotic behavior of $V$ as $\lambda$
goes to infinity.

\begin{proposition}
	\label{Convergencia puntual con lambda a infinito} For all
	$x\geq0$, it holds that $\lim_{\lambda\rightarrow\infty}V(x,\lambda)=x$.
\end{proposition}

\section{HJB equation and viscosity solutions for the auxiliary
	problem\label{Seccion Viscosidad}}
In this section we obtain the Hamilton-Jacobi-Bellman equation
associated to the auxiliary problem and state that the optimal value function
$V$ defined in (\ref{DefinitionV}) is a viscosity solution. We also show that
the value function of any admissible strategy is smaller than any
supersolution of the HJB equation. Finally, we characterize the optimal value
function $V$ as the smallest viscosity supersolution of the HJB equation.
\ These results are a generalization of the ones given in {\cite[Sec.4]{AAM24}} in the absence of a switching time. The proofs of
the results of this section are deferred to Appendix
\ref{Appendix Section HJB}.

The HJB equation of the optimization problem \eqref{DefinitionV} is given by%

\begin{equation}
	\max\{\mathcal{L}(V)(x,\lambda),1-V_{x}(x,\lambda)\}=0,\label{HJB}%
\end{equation}
where%

\begin{equation}%
	\begin{array}
		[c]{lll}%
		\mathcal{L}(V)(x,\lambda) & = & pV_{x}(x,\lambda)-d\left(  \lambda
		-\underline{\lambda}\right)  V_{\lambda}(x,\lambda)-(q+\lambda+\beta
		)V(x,\lambda)\\
		&  & +\lambda%
		{\textstyle\int\nolimits_{0}^{x}}
		V(x-\alpha,\lambda)dF_{U}(\alpha)+\beta%
		{\textstyle\int\nolimits_{0}^{\infty}}
		V(x,\lambda+\gamma)dF_{Y}(\gamma)\\
		&  & +{\small \xi}\left(  {\small g(x,\lambda)-V(x,\lambda)}\right)  .
	\end{array}
	\label{DefinicionL}%
\end{equation}
{While the derivation of Equation \eqref{HJB} follows standard procedures, it may be
	instructive to interpret it concretely. At each point in time, one has to
	decide, based on the current surplus level $x$ and intensity level $\lambda $%
	, whether paying dividends is preferable to not paying them. The left-hand
	side in the max operator refers to the case of no dividend payment. In that
	scenario, the process evolves according to the infinitesimal generator given
	in \eqref{DefinicionL}. As the premium rate is $p$, the infinitesimal change in the $x$-direction in the absence of a claim is given by $pV_{x}(x,\lambda )$. A
	claim arrives with intensity $\lambda $ so we might interpret this as the
	surplus process being 'killed' at that rate (term $-\lambda V(x,\lambda )$)
	and immediately 'revived' with a modified surplus level that depends on the
	claim size with c.d.f. $F_{U}$ (integral term $\lambda%
	{\textstyle\int\nolimits_{0}^{x}}
	V(x-\alpha,\lambda)dF_{U}(\alpha)$). In the $\lambda $-direction, in the absence of a jump in the
	shot-noise process, the infinitesimal change in the claim intensity level is
	described by the term $-d(\lambda -\underline{\lambda })V_{\lambda
	}(x,\lambda )$ (which, due to the Markov property of the intensity process,
	also comprises the decay of all previous intensity shots), and since an intensity
	shot arrives with intensity $\beta $ we might interpret this as the
	process being 'killed' ( $-\beta V(x,\lambda )$) and 'revived'
	again with the respective up-jump with c.d.f. $F_{Y}$ (which is captured in
	the integral term $\beta%
	{\textstyle\int\nolimits_{0}^{\infty}}
	V(x,\lambda+\gamma)dF_{Y}(\gamma)$). Finally, the term $-qV(x,\lambda )$
	accounts for the time value due to the discounting of future payments,
	and the term $\xi (g(x,\lambda )-V(x,\lambda ))$ keeps track of switching
	events, which occur with intensity $\xi $ and replace $V(x,\lambda )$ with
	its boundary value $g(x,\lambda )$ upon switching. Note that $\xi =0$ leads
	to the integro-differential operator considered in \cite{AAM24} without switching
	time $\overline{\tau }$. Moreover, one can interpret that it is not optimal
	to pay out dividends if $V_{x}(x,\lambda )$ is larger than $1$, because
	paying out dividends at that moment will realize a slope of $1$ in the
	direction of the surplus $x$ . This means that one benefits from
	the dynamics of the surplus process more than when paying dividends,
	translating into the right-hand side of \eqref{HJB} being negative in that case.}\\
\noindent {The structure of Equation \eqref{HJB} suggests that the state space
$[0,\infty)\times\lbrack\underline{\lambda},\infty)$ is partitioned into a no-action region $\mathcal{NA}$ (in which no dividends are paid) and
an action region $\mathcal{A}$ (in which dividends are paid). The points in the $\mathcal{NA}$ region satisfy $\mathcal{L}(V)=0$ and
$1-V_{x}$ $<0$, whereas the points in $\mathcal{A}$ satisfy $\mathcal{L}(V)\leq0$
and $1-V_{x}$ $=0$.}\\

\noindent Due to Proposition \ref{Basic Results V}, the optimal value function $V$ is locally Lipschitz. However, it might not be differentiable at
some points, so we cannot claim that $V$ is a solution of the HJB equation in
the classical sense. We prove instead that $V$ is a viscosity solution of the HJB equation. Let us define this standard notion.

\begin{definition}
	\label{Viscosity}A uniformly continuous function $\overline{u}:\left[
	0,\infty\right)  \times\left[  \underline{\lambda},\infty\right)
	\rightarrow{\mathbb R}$\textbf{,} that is locally Lipschitz\ in $(0,\infty
	)\times\left[  \underline{\lambda},\infty\right)  $, is a \textit{viscosity
		supersolution} of (\ref{HJB})\ at $(x,\lambda)\in(0,\infty)\times
	(\underline{\lambda},\infty)$\ if any continuously differentiable function
	$\varphi:(0,\infty)\times\left[  \underline{\lambda},\infty\right)
	\rightarrow{\mathbb R}\ $such that $\varphi\left(  y,\cdot\right)  $ is bounded
	for any $y\geq0,~\varphi(x,\lambda)=\overline{u}(x,\lambda)$ and such that
	$\overline{u}-\varphi$\ reaches the minimum at $\left(  x,\lambda\right)
	$\ satisfies
	\[
	\max\left\{  \mathcal{L}(\varphi)(x,\lambda),1-\varphi_{x}(x,\lambda)\right\}
	\leq0.\
	\]
	A uniformly continuous function $\underline{u}:\left[  0,\infty\right)
	\times\left[  \underline{\lambda},\infty\right)  \rightarrow{\mathbb R}$, that
	is locally Lipschitz\ in $(0,\infty)\times\left[  \underline{\lambda}%
	,\infty\right)  $, is a \textit{viscosity subsolution} of (\ref{HJB})\ at
	$(x,\lambda)\in(0,\infty)\times(\underline{\lambda},\infty)$\ if any
	continuously differentiable function $\ \psi:(0,\infty)\times
	(\underline{\lambda},\infty)\rightarrow{\mathbb R}\ $such that $\psi\left(
	y,\cdot\right)  $ is bounded for any $y\geq0$, $\psi(x,\lambda)=\underline{u}%
	(x,\lambda)$ and such that $\underline{u}-\psi$\ reaches the maximum at
	$\left(  x,\lambda\right)  $ satisfies
	\[
	\max\left\{  \mathcal{L}(\psi)(x,\lambda),1-\psi_{x}(x,\lambda)\right\}
	\geq0\text{.}%
	\]
	A function $u:(0,\infty)\times(\underline{\lambda},\infty)\rightarrow
	{\mathbb R}$ which is both a supersolution and subsolution at $(x,\lambda
	)\in(0,\infty)\times(\underline{\lambda},\infty)$ is called a
	\textit{viscosity solution} of (\ref{HJB})\ at $\left(  x,\lambda\right)  $.
\end{definition}

\begin{proposition}
	\label{Prop V is a viscosity solution} $V$ is a viscosity solution of the HJB
	equation (\ref{HJB}) at any $\left(  x,\lambda\right)  \in(0,\infty
	)\times(\underline{\lambda},\infty)$.
\end{proposition}

\noindent The optimal value function can now be characterized in the following way.

\begin{proposition}
	\smallskip\label{MenorSuper} The optimal value function $V$ is the smallest
	viscosity supersolution of (\ref{HJB}) among those that are non-increasing in
	$\lambda$ and satisfy the growth condition (\ref{Growth Condition}).
\end{proposition}

\noindent Next, the usual viscosity verification result can be deduced.

\begin{corollary}
	\label{VerificacionDividendos} Consider a family of admissible strategies
	$\{L^{x,\lambda}\in\Pi_{x,\lambda}:\left(  x,\lambda\right)  \in
	(0,\infty)\times(\underline{\lambda},\infty)\}$. If the function
	$W(x,\lambda):=J(L^{x,\lambda};x,\lambda)$ is a viscosity supersolution of
	(\ref{HJB}), then $W(x,\lambda)$ is the optimal value function. Also, if for
	each $k\geq1$ there exists a family of strategies $\{L_{k}^{x,\lambda}\in
	\Pi_{x,\lambda}:\left(  x,\lambda\right)  \in(0,\infty)\times
	(\underline{\lambda},\infty)\}$ such that $W(x,\lambda):=\lim_{k\rightarrow
		\infty}J(L_{k}^{x,\lambda};x,\lambda)$ is a viscosity supersolution of the HJB
	equation (\ref{HJB}) in $(0,\infty)\times(\underline{\lambda},\infty)$, then
	$W$ is the optimal value function $V$.
\end{corollary}

\section{Asymptotic properties of the optimal value function for the auxiliary
	problem\label{Section Asymptotic}}

From Corollary \ref{VerificacionDividendos} we get the next proposition, in
which we prove that there exists an explicit threshold $\overline{x}$, such
that if the surplus is above of this explicit threshold, then one should pay
at least the exceeding surplus as dividends, regardless of the current
intensity of the shot-noise process; so $[\overline{x},\infty)\times
\lbrack\underline{\lambda},\infty)\subset$ $\mathcal{A}$ . However, as we will
show in the examples, the boundaries between the no-action region
$\mathcal{NA}$ and the action region $\mathcal{A}$ depend on $\lambda$, so it {is} a complex problem and the value $\overline{x}$ is just an upper bound. The
proofs of the results of this section and some needed auxiliary results are given in Appendix \ref{Appendix Asymptotic}.

\begin{proposition}
	\label{Proposicion acotacion de retirar} $V(x,\lambda)=x-\overline
	{x}+V(\overline{x},\lambda)$ for $x\geq\overline{x}$, where%
	\[
	\overline{x}:=\max\left\{  \frac{p+{\small \xi}\left(  {\small g(}%
		\widehat{x}{\small ,\lambda)-}\widehat{x}\right)  }{q},\widehat{x}\right\}
	\]
	{and $\hat{x}$ is the one defined in Assumption (A1).} 
\end{proposition}

\indent We have the following uniform convergence as the current intensity of the
shot-noise process goes to infinity.

\begin{proposition}
	\label{Convergencia Uniforme con Lambda} $\lim_{\lambda\rightarrow\infty
	}\left(  \sup_{x\geq0}V(x,\lambda)-x\right)  =0.$
\end{proposition}

\section{Numerical scheme for the auxiliary problem \label{Numerical Scheme}}

We describe here the numerical method that is based on admissible strategies
restricted to a finite grid. The numerical scheme is derived from the one in \cite{AM 2021} and it is a generalization of the one
described in Sections 6, 7 and 8 in {\cite{AAM24}} when there is
no-switching time $\overline{\tau}$ (i.e., $\xi=0$). Proposition
\ref{Convergencia Uniforme con Lambda} suggests that we can approximate the
optimal value function {by the one over} a bounded grid on $\lambda$. Also, from
Proposition \ref{Proposicion acotacion de retirar}, we can assume that
$(x,\lambda)\in\mathcal{A}$ for $x>\overline{x}\ $. This allows to make a
numerical computation staying in a finite grid. {For some} $\delta>0$, $\Delta>0 $
and $m_{1}\in\mathbb{N}$ , we define the finite grid%
\[
\overline{\mathcal{G}}_{\delta}\times\overline{\mathcal{H}}_{\Delta},
\]
where%
\[
\overline{\mathcal{G}}_{\delta}=\left\{  x_{n}^{\delta}=np\delta:x_{n}%
^{\delta}\leq\overline{x}\ \right\}  \;\text{and}\;\;\overline{\mathcal{H}}_{\Delta
}=\left\{  \lambda_{m}^{\Delta}=\underline{\lambda}+m\Delta:0\leq m\leq
m_{1}\right\}.
\]
Let us call $n_{1}=\max\left\{  n:x_{n}^{\delta}\in\overline{\mathcal{G}%
}_{\delta}\right\}$. In order to discretize on $\lambda$, we consider the
auxiliary process$\ $%
\[
\widehat{\lambda}_{t}:=\sigma^{\Delta}\left(  \lambda_{t}\right)
\wedge\lambda_{m_{1}}^{\Delta}\in\overline{\mathcal{H}}_{\Delta}%
\]
instead of the original shot-noise process $\boldsymbol{\lambda}=\left(
\lambda_{t}\right)  _{t\geq0}$, where%

\begin{equation}
	\sigma^{\Delta}\left(  \lambda\right)  =\min\left\{  \lambda_{m}^{\Delta}%
	\in\mathcal{H}_{\Delta}:\lambda_{m}^{\Delta}\geq\lambda\right\}
	\in\mathcal{H}_{\Delta}.\label{Definicion Lambda en grilla}%
\end{equation}
So, if $\lambda_{t}\leq\lambda_{m_{1}}^{\Delta}$, we are discretizing
$\lambda_{t}$ considering the closest largest point in $\mathcal{H}_{\Delta}$, 
and if $\lambda_{t}\geq\lambda_{m_{1}}^{\Delta}$ then $\widehat{\lambda}%
_{t}=\lambda_{m_{1}}^{\Delta}$. We also introduce the auxiliary function
\begin{equation}
	\rho^{\delta}(x):=\max\{x_{n}^{\delta}\in\overline{\mathcal{G}}:x_{n}^{\delta
	}\leq x\},\label{Definicion de x en la grilla  por abajo}%
\end{equation}
$\ $which gives the closest point of the grid $\overline{\mathcal{G}}_{\delta
}$ below $x.$

Define now the following local strategies: either the company pays no dividends or
pays immediately a lump sum $p\delta$ as dividends; moreover, the company can
finish the insurance contract at any time. We modify these local strategies in
such a way that the controlled surplus and the intensity always lie in
$\overline{\mathcal{G}}_{\delta}\times\overline{\mathcal{H}}_{\Delta}$
immediately after the arrival of a claim. Let $\tau^U$ and $U$ be the arrival
time and the size of the next claim, and $T$ and $Y$ be the arrival time and
the size of the next intensity upward jump. We first define the three possible
control actions at any point of $\left(  x_{n}^{\delta},\lambda_{m}^{\Delta
}\right)  \in$ $\overline{\mathcal{G}}_{\delta}\times\overline{\mathcal{H}%
}_{\Delta}$ as follows:

\begin{itemize}
	\item Control action $\mathbf{E}_{0}$: Pay no dividends up to the time
	$\delta\wedge\tau^U\wedge T\wedge\overline{\tau}$. The control action
	$\mathbf{E}_{0}$\textbf{\ }can only be applied for current surplus with
	$n<n_{1}$.
	
	\begin{enumerate}
		\item In the case that $\delta<\tau^U\wedge T\wedge\overline{\tau}$, the surplus
		at time $\delta$ is $x_{n+1}^{\delta}\in\overline{\mathcal{G}}_{\delta}$, and
		the intensity is $\widehat{\lambda}_{\delta}=\sigma^{\Delta}\left(
		\underline{\lambda}+e^{-d\delta}\left(  \lambda_{m}^{\Delta}%
		-\underline{\lambda}\right)  \right)  .$
		
		\item If $\tau^U<\delta\wedge T\wedge\overline{\tau}$, the uncontrolled surplus
		at time $\tau^U$ is $x_{n}^{\delta}+\tau^U p-U$ and the intensity is
		$\widehat{\lambda}_{\tau^U}=\sigma^{\Delta}\left(  \underline{\lambda}%
		+e^{-d\tau^U}\left(  \lambda_{m}^{\Delta}-\underline{\lambda}\right)\right);$ if $x_{n}^{\delta}+\tau^U p-U$ is positive, the company pays immediately the
		minimum amount of dividends in such a way that the end surplus is
		$\rho^{\delta}(x_{n}^{\delta}+\tau^U p-U)$ and the amount paid as dividends is
		equal to $x_{n}^{\delta}+\tau^U p-U-\rho^{\delta}(x_{n}^{\delta}+\tau^U p-U)$. Ruin occurs in the case that the surplus $x_{n}^{\delta}+\tau^U p-U<0$ at time
		$\tau^U\leq$ $\delta\wedge T\wedge\overline{\tau}$.
		
		\item If $T<\delta\wedge\tau^U\wedge\overline{\tau},$ the end surplus is
		$\rho^{\delta}(x_{n}^{\delta}+Tp)$, the amount paid as dividends is
		$x_{n}^{\delta}+Tp-\rho^{\delta}(x_{n}^{\delta}+Tp)$ and the intensity is
		$\widehat{\lambda}_{T}=\sigma^{\Delta}\left(  \underline{\lambda}%
		+e^{-dT}\left(  \lambda_{m}^{\Delta}-\underline{\lambda}\right)  +Y\right)  .
		$
		
		\item If $\overline{\tau}<T\wedge\delta\wedge\tau^U$, the compound Cox process
		switches and the exit value of the company becomes $g(x_{n}^{\delta}%
		+\overline{\tau}p,\widehat{\lambda}_{\overline{\tau}})$, where
		$\widehat{\lambda}_{\overline{\tau}}=\sigma^{\Delta}\left(  \underline{\lambda
		}+e^{-d\overline{\tau}}\left(  \lambda_{m}^{\Delta}-\underline{\lambda
		}\right)  \right)  $.
	\end{enumerate}
	
	\item Control actions $\mathbf{E}_{1}$: The company pays immediately $p\delta$
	as dividends, so the controlled surplus becomes $x_{n-1}^{\delta}%
	\in\mathcal{G}_{\delta}$. The control action $\mathbf{E}_{1}$\textbf{\ }can
	only be applied for current surplus $x_{n}^{\delta}>0$.
	
	\item Control action $\mathbf{E}_{F}$: The company opts to pay the current
	surplus $x_{n}^{\delta}$ as dividends and {immediately terminates the insurance policies}.
\end{itemize}

\noindent We denote the space of control actions as
\begin{equation}
	\mathcal{E}=\{\mathbf{E}_{F},\mathbf{E}_{1},\mathbf{E}_{0}%
	\}.\label{Conjunto E de Estrategias.}%
\end{equation}
\noindent The numerical scheme will choose inductively which of the possible local actions in each
of the points in the grid is "better": $\mathbf{E}_{0}$ or $\mathbf{E}_{1}$ or
$\mathbf{E}_{F}$. Roughly speaking, the idea of this construction is to find,
for each point in $\overline{\mathcal{G}}_{\delta}\times\overline{\mathcal{H}%
}_{\Delta}$, the best local strategy (in the sense that maximizes expected
discounted dividends until the switching random exponential time
$\overline{\tau}$, after which the terminal value is $g$) restricted to the
family of control actions defined above. For that purpose, let us introduce
the operators related to the control actions in $\mathcal{E} $. Consider the
operators $\widehat{\mathcal{T}}_{0}$, $\mathcal{T}_{1}$ and $\mathcal{T}_{F}$
in the set of functions
\[
\mathcal{W}^{\delta,\Delta}=\left\{  w:\mathcal{G}_{\delta}\times
\mathcal{H}_{\Delta}\rightarrow\lbrack0,\infty)\right\}
\]
defined as follows%

\begin{equation}
	\widehat{\mathcal{T}}_{0}(w)(x_{n}^{\delta},\lambda):=\mathbb{P}%
	(\delta<T\wedge\tau^U\wedge\overline{\tau})e^{-q\delta}w(x_{n+1}^{\delta
	},\widehat{\lambda}_{\delta})+\widehat{\mathcal{I}}(w)(x_{n}^{\delta}%
	,\lambda_{m}^{\Delta}),\label{Definicion TE}%
\end{equation}
where%

\[%
\begin{array}
	[c]{l}%
	\widehat{\mathcal{I}}(w)(x_{n}^{\delta},\lambda_{m}^{\Delta})\\%
	\begin{array}
		[c]{l}%
		:=\mathbb{E}(I_{\delta\wedge T\wedge\tau^U\wedge\overline{\tau}=\tau^U}I_{\left\{
			x+p\tau^U-U\geq0\right\}  }e^{-q\tau^U}w(\rho^{\delta}(x_{n}^{\delta}%
		+p\tau^U-U),\widehat{\lambda}_{\tau^U}))\\
		+\mathbb{E}(I_{\delta\wedge T\wedge\tau^U\wedge\overline{\tau}=T}~e^{-qT}%
		w(x_{n}^{\delta},\widehat{\lambda}_{T}))+\mathbb{E}(I_{\delta\wedge
			T\wedge\tau^U\wedge\overline{\tau}=T}e^{-qT}pT)\\
		+\mathbb{E}(I_{\delta\wedge T\wedge\tau^U\wedge\overline{\tau}=\tau^U}I_{\left\{
			x+p\tau^U-U\geq0\right\}  }e^{-q\tau^U}(x_{n}^{\delta}+p\tau^U-U-\rho^{\delta}%
		(x_{n}^{\delta}+p\tau^U-U)))\\
		+\mathbb{E}(I_{\delta\wedge T\wedge\tau^U\wedge\overline{\tau}=\overline{\tau}%
		}~e^{-q\overline{\tau}}g(x_{n}^{\delta}+p\overline{\tau},\widehat{\lambda
		}_{\overline{\tau}}));
	\end{array}
\end{array}
\]%
\begin{equation}%
	\begin{array}
		[c]{ccc}%
		\mathcal{T}_{1}(w)(x_{n}^{\delta},\lambda):=w(x_{n-1}^{\delta},\lambda)+\delta
		p & \text{and} & \mathcal{T}_{F}(w)(x_{n}^{\delta},\lambda):=x_{n}^{\delta}.
	\end{array}
	\label{Definicion Ti TS}%
\end{equation}
Here $\tau^U$ and $U$ are the arrival time and the size of the next claim, and $T$
and $Y$ are the arrival time and the size of the next intensity upward jump. We
also introduce the operator $\widehat{\mathcal{T}}$ in $\mathcal{W}^{\delta}$
as%
\begin{equation}
	\widehat{\mathcal{T}}:=\max\{\widehat{\mathcal{T}}_{0},\mathcal{T}%
	_{1},\mathcal{T}_{F}\}.\label{Definicion T}%
\end{equation}
The operator $\widehat{\mathcal{T}}_{0}$ is associated to the local action
$\mathbf{E}_{0}$, the operator $\mathcal{T}_{1}$ to the local action
$\mathbf{E}_{1}$ and the operator $\mathcal{T}_{F}$ to the local action
$\mathbf{E}_{F}.$ The control action $\mathbf{E}_{F}$ is never optimal but it
facilitates the numerical scheme. It can be proved, much like in {\cite{AAM24}}, that there exists a unique fixed point of $\widehat{\mathcal{T}}$
which approximates uniformly the optimal value $V$ taking $\delta$ and
$\Delta$ small enough and $m_{1}$ large enough. Let us denote this unique fixed
point of $\widehat{\mathcal{T}}$ as $\widehat{V}^{\delta,\Delta}$, so
$\widehat{\mathcal{T}}(\widehat{V}^{\delta,\Delta})=\widehat{V}^{\delta
	,\Delta}.$ We can then state the following result (the proofs of which are totally analogous to the ones of Proposition 7.2 and Theorem 7.4 in {\cite{AAM24}}, so we omit them here). 

\begin{proposition}
	\label{Teorema Aproximacion Uniforme}For any $\varepsilon>0$, there exist
	$\delta$ and $\Delta$ small enough so that $0\leq V-\widehat{V}^{\delta,\Delta}%
	\leq$ $\varepsilon$ in $[0,\infty)\times\lbrack\underline{\lambda},\infty).$
\end{proposition}

\noindent In order to obtain an approximation of $\widehat{V}^{\delta,\Delta}$, we
proceed inductively in the following \ way:

\begin{enumerate}
	\item We define as the initial function $\widehat{W}_{1}(x_{n}^{\delta
	},\lambda_{m}^{\Delta})=x_{n}^{\delta}.$ Note that $\widehat{W}_{1}$
	corresponds to applying the local action $\mathbf{E}_{F}$ in all the points of
	the grid.
	
	\item We calculate $\widehat{W}_{l+1}=\widehat{\mathcal{T}}(\widehat{W}_{l})$. The action regions $\widehat{\mathcal{A}}_{l+1}$ {(where dividends are paid)} and no-action regions
	$\widehat{\mathcal{NA}}_{l+1}${(where no dividends are paid)} associated to $\widehat{W}_{l+1}$ are obtained
	as%
	\[%
	\begin{array}
		[c]{lll}%
		\widehat{\mathcal{A}}_{l+1} & = & \{(x_{n}^{\delta},\lambda_{m}^{\Delta}%
		)\in\overline{\mathcal{G}}_{\delta}\times\overline{\mathcal{H}}_{\Delta
		}:\widehat{\mathcal{T}}(\widehat{W}_{l})(x_{n}^{\delta},\lambda_{m}^{\Delta
		})=\widehat{\mathcal{T}}_{1}(\widehat{W}_{l})(x_{n}^{\delta},\lambda
		_{m}^{\Delta})\},\\
		\widehat{\mathcal{NA}}_{l+1} & = & \{(x_{n}^{\delta},\lambda_{m}^{\Delta}%
		)\in\overline{\mathcal{G}}_{\delta}\times\overline{\mathcal{H}}_{\Delta
		}-\widehat{\mathcal{A}}_{l+1}:\widehat{\mathcal{T}}(\widehat{W}_{l}%
		)(x_{n}^{\delta},\lambda_{m}^{\Delta})=\widehat{\mathcal{T}}_{0}%
		(\widehat{W}_{l})(x_{n}^{\delta},\lambda_{m}^{\Delta})\}.
	\end{array}
	\]
	Note that $\mathbf{E}_{F}$ is never optimal, we have just used $\mathbf{E}%
	_{F}$ to construct the initial function $\widehat{W}_{1}.$
\end{enumerate}

The proof that $\lim_{l\rightarrow\infty}\widehat{W}_{l}$ $=\widehat{V}%
^{\delta,\Delta}$ is as in Section 8 in {\cite{AAM24}}.

\section{A compound shot-noise risk model with a tipping point\label{Climate Tipping Point}}
The auxiliary risk model studied in the previous sections can now be used to study a full model with a tipping point $\omega$, where the surplus process of an insurance company {evolves} according to a compound Cox shot-noise risk model as defined in Section \ref{Model and basic results} with a certain set of parameters and distributional assumptions up to a random time horizon $\omega$ (not necessarily exponentially distributed). After time $\omega$, the compound Cox process switches to
another compound Cox process. Concretely, assume that we have a random tipping point $\omega\sim \text{Erlang}(k,\xi)$ for a given $k$ and $\xi$. Then \[
\omega=\omega^{k}+\omega^{k-1}+....+\omega^{1},%
\]
where $\omega^{i}$\ are independent i.i.d.\ exponential random variables with
parameter $\xi$. \\
{A main motivation for the choice of an Erlang distribution for $\omega$ is tractability, as it exploits the memoryless property iteratively, and yet allows to deviate from an exponential shape of the respective density of the waiting time to much more peaked scenarios, as $k$ grows (and $\xi$ is chosen to,  for instance, match a certain expected value). An even more general choice would be the class of matrix-exponential distributions, which is dense in the class of all distributions on the positive half-line (see e.g. \cite{BladtNielsen}). However, for the purposes of the present paper, there is no added value to go to that considerably higher level of complexity in the calculations. \\
In terms of concrete parameter estimation, if one had sufficiently many historical instances for tipping point occurrences in the past, one could then develop statistical procedures to estimate $k$ and $\xi$ for $\omega\sim \text{Erlang}(k,\xi)$ with respect to some optimization criteria. However, historical data for such events are scarce, and for the present purpose we would suggest to estimate the expected waiting time $k/\xi$ (for instance according to some concrete climate projections, cf.\ \cite{LentonTipping}) and the degree of uncertainty around that value being represented by the value of $k$. Furthermore, we assume here that all 
$\omega^{i}$ are observable. That may be a realistic assumption in situations where the approaching of the tipping point is a gradual process with some climate indicators reaching intermediate thresholds, and the arrival of the tipping point $\omega$ itself may also be considered as a major observable instance. In case that $\omega^{i}$ is not observable, one faces a filtering problem for the state $i$, which is an additional complication, but may still lead to solvable cases, see e.g.\ \cite{filter} for another filtering context.}\\

We will now illustrate how to use the auxiliary problem (iteratively and
{in a backward direction}) to solve the optimal dividend problem for this case. \\
We call State $m$, where $1\leq m\leq k,$ the
state in which the time until the tipping point will happen is given
by $\omega^{m}+\omega^{m-1}+....+\omega^{1}$. Let us call State $0$ the
final compound Cox process, after the tipping point happened. Hence, starting at
State $m$, at a random exponential time $\omega^{m} $\ the process switches to
State $m-1$. To solve the problem of maximizing dividends, we therefore need to {apply} an iterative backward procedure: from the final State $0$ to State $1$ to
State $2$ and so on.

Let us first consider State $0$ (that is after the tipping point). Given an
initial surplus $x$ and initial intensity $\lambda,$ the expected discounted
dividends until ruin for an admissible strategy $L=(L_{t})_{t\geq0}$ \ {are}
given by
\[
J^{0}(L)={\mathbb E}\left[  \int_{0}^{\tau}e^{-qs}dL_{s}\right]  ,
\]
where $q>0$\ is the discount rate and $\tau$ the ruin time. The optimal
value function is given by
\[
V^{0}(x,\lambda)=\sup_{L}J^{0}(L).
\]
This problem is exactly the one considered in {\cite{AAM24}}, and it 
can be solved with the same numerical procedure described in the previous
section, but putting $\xi=0$. Consider now State $m\geq1$. Let us define the
optimal value function $V^{m}$ in State $m$, assuming that we have already obtained $V^{m-1}$. This problem can be solved using the auxiliary problem
defined in (\ref{DefinitionV}) and (\ref{Definicion VL}) calling $V:=V^{m}$
and $g:=V^{m-1}$. Before and after the climate tipping point, the compound Cox
process can be completely different, the unique restriction is that the base-line
intensities fulfill $\underline{\lambda}_{m}\geq\underline{\lambda}_{m-1}$.\\

The same technique can in fact be used for more than one climate tipping point.
Consider for instance two consecutive tipping points: the first one arriving at an 
Erlang$(k_{1},\xi_{1})$ time, and the second one at an Erlang$(k_{2},\xi_{2})$ time later. One can then obtain numerically the value functions $V^{0},V^{1},...,V^{k_{2}}$ for the
last tipping point as described before, then define
the auxiliary problem with $g:=V^{k_{2}}$ and start again solving the problem
iteratively and backwardly for the tipping point Erlang$(k_{1},\xi_{1})$ as
before. In that case, one would hence need to solve $k_{1}+k_{2}+1$ numerical
auxiliary problems. The short-coming of the latter procedure is that despite the fact that in each step
the numerical approximation of the value function converges to the optimal
value function, the numerical errors may accumulate as the number of steps increase.

\section{Numerical Results \label{Numerical Results}}

In this section, we show some numerical results. Due to the uniform convergence result, when $\delta$
and $\Delta$ are sufficiently small, the optimal value function can be obtained with an arbitrarily small approximation error. In our numerical scheme we ensure that $\delta~$and $\Delta$ are sufficiently small by
comparing the value functions for $(\delta,\Delta)$ with the ones of
$(\delta/2,\Delta/2)$. We consider examples with varying claim
size distributions and upward jump distributions in the shot-noise intensity process.\\

{We will first consider an example with exponential claim sizes, where at an Erlang(2)-distributed tipping point the intensity of catastrophe arrivals increases. It turns out that in this situation a barrier strategy is optimal where the barrier level depends on both current surplus level $x$ and intensity level $\lambda$. In Section \ref{sec7.2} we then consider an example with Erlang(2) claim sizes for a situation where under constant intensity a two-band strategy is known to be optimal, and we show that for certain $\lambda$-values such a two-band remains optimal, whereas for small and large values of $\lambda$ it collapses to a barrier strategy, where the barrier level again depends on $x$ and $\lambda$. In Section \ref{sec7.3} we then consider a situation where at the tipping point not only the catastrophe arrival dynamics, but also the claim size distribution changes. Finally, in Section \ref{sec7.4} we consider a variant of the example in Section \ref{sec7.3} with truncated claims due to a policy limit.}

\subsection{Example 1: Exponential claims and increasing number of catastrophies}

Consider a climate tipping point at an Erlang$\left(  2,1/3\right)$ time {(so the expected time until its arrival is 6 years)} with the
frequency of catastrophes increasing after the tipping point. The base
intensity equals $\underline{\lambda}=1/4$, and we assume an exponential intensity jump
distribution with cdf $F_{Y}(x)=1-e^{-x/2}$, $d=7/10$ (so that the additional
claim arrival intensity due to a catastrophe is halved after one year), as well as an
exponential claim size distribution with cdf $F_{U}(x)=1-e^{-10x}$. The discount
rate is chosen to be $q=1/5$ and the insurance safety loading applied in the policies equals
$\eta=1/5$. {The concrete values chosen are of realistic magnitudes, and they are also chosen in such a way so that the action and non-action regions in the plots are nicely visible}. Before the tipping point we consider the intensity of jump
intensity arrivals to be $\beta=1/3$ (so we expect a catastrophe
to happen every three years) and after the tipping point we assume $\beta=1/2$ (i.e., afterwards we expect a catastrophe to happen every two years). We then obtain from
(\ref{Definicion p}), that the premium rate equals $p=101/700$ before the
tipping point and $p=141/700$ after the tipping point. Note that we have designed the example in such a way that the setting after the tipping point is exactly the one of Example 1 of \cite[Sec.8]{AAM24}.
For the grid parameters, it turns out that to consider a finite grid of
$60\times60$ points in the compact set $(x,\lambda)\in\left[  0,0.4\right]
\times\left[  1/4,5\right]  $ is sufficient here. All the points outside this compact set are defined to be in the action region. 

\begin{figure}[ptb]
	\begin{subfigure}{.32\textwidth}
		\centering
		\includegraphics[width=5.5cm]{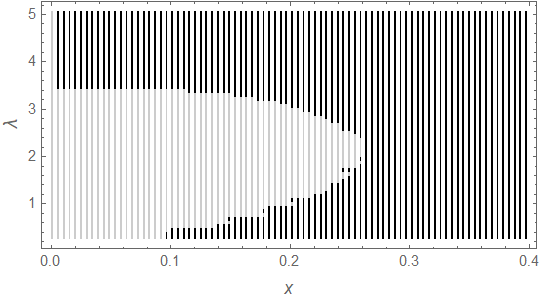}
		\caption{State 2}
		\label{Ex1_1}
	\end{subfigure} \begin{subfigure}{.32\textwidth}
		\centering
		\includegraphics[width=5.5cm]{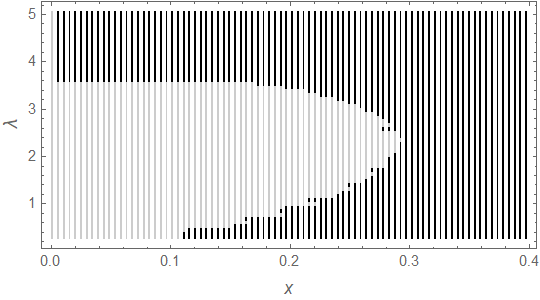}
		\caption{State 1}
		\label{Ex1_2}
	\end{subfigure}
	\begin{subfigure}{.32\textwidth}
		\centering
		\includegraphics[width=5.5cm]{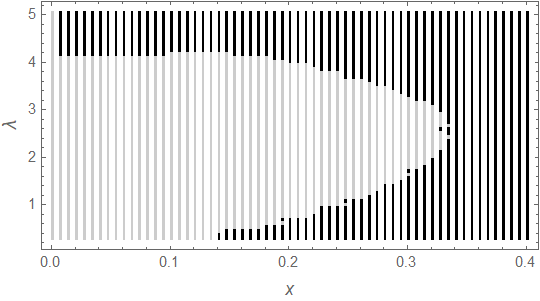}
		\caption{State 0}
		\label{Ex1_3}
	\end{subfigure}
	\caption{{\small Action regions (dark) and no-action regions (light) of the optimal strategy as a function of $\lambda$ and $x$ for each of the states in Example 1}}%
	\label{fig1}%
\end{figure}

Figure \ref{fig1} depicts the action (dark) and no-action
(light) regions of the optimal dividend strategy in the different states as a function of current surplus level $x $ and
current intensity value $\lambda$. In Figure \ref{Ex1_1} the regions are plotted for State 2,
which is the situation in the beginning ($t=0$). Figure \ref{fig1}b then shows the regions in State 1, when the first of the two exponential components of the Erlang(2) waiting time has already materialized. One sees that if (for some reason) State 1 is observable (i.e.\ one knows to be already closer to the tipping point, due to environmental variables having deteriorated considerably, temperature having risen further etc.), then the dividend strategy should already be adapted, namely the dividend barrier (as a function of current intensity $\lambda$) should be raised. Finally
Fig \ref{fig1}c depicts the optimal action and no-action regions after the tipping point (State 0), which leads to a strategy where less dividends are paid (and this plot is identical to the one of \cite[Fig.8.1(A)]{AAM24}). 

\begin{figure}[ptb]
	\begin{subfigure}{.32\textwidth}
		\centering
		\includegraphics[width=4.5cm]{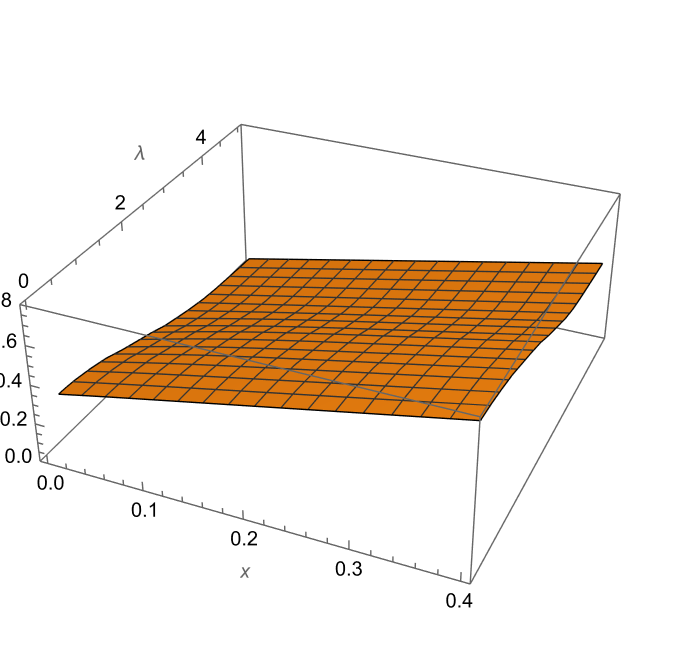}
		\caption{$V^2(x,\lambda)$}
		\label{Ex1_val2}
	\end{subfigure} \begin{subfigure}{.32\textwidth}
		\centering
		\includegraphics[width=4.5cm]{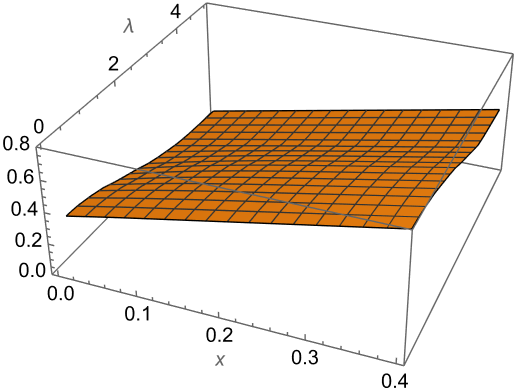}
		\caption{$V^1(x,\lambda)$}
		\label{Ex1_val1}
	\end{subfigure}
	\begin{subfigure}{.32\textwidth}
		\centering
		\includegraphics[width=5.5cm]{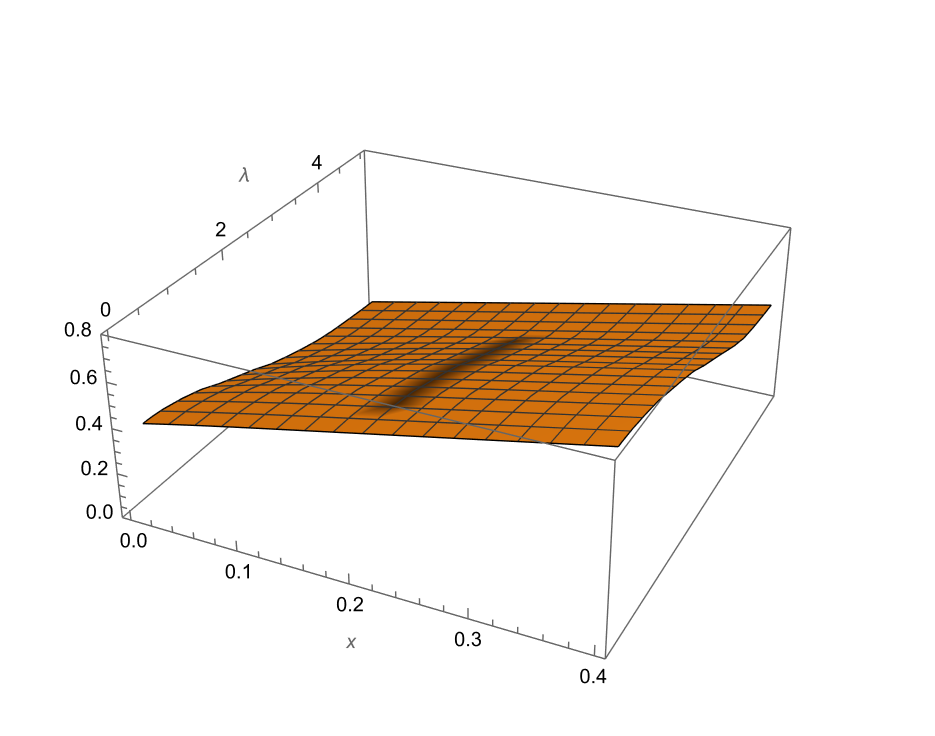}
		\caption{$V^0(x,\lambda)$}
		\label{Ex1_val0}
	\end{subfigure}
	\caption{\small Value functions $V^i(x,\lambda)$ {as a function of $x$ and $\lambda$} for each State $i$ in Example 1}%
	\label{fig2}%
\end{figure}

The optimal value functions $V^{2}$, $V^{1}$ and $V^{0}$ in State
2, State 1 and State 0, respectively, are depicted in Figure \ref{fig2} as a function of initial surplus $x$ and initial claim
 intensity level $\lambda$ (Figure \ref{Ex1_val0} again corresponds to \cite[Fig.8.1(B)]{AAM24}). \\

\begin{figure}[ptb]
		\centering
 \begin{subfigure}{.32\textwidth}
	\centering
	\includegraphics[width=5.5cm]{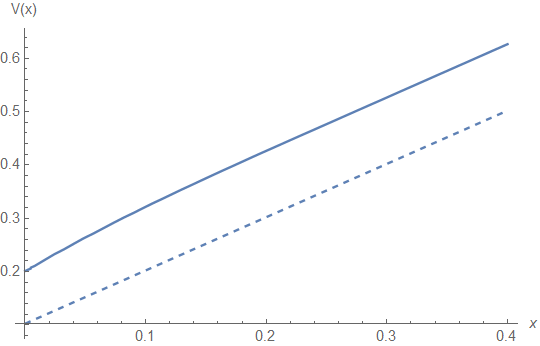}
	\caption{$V^2(x,\lambda_{\text{av}})$ vs. $V_{\text{CL}}(x,\lambda_{\text{av}})$}
	\label{Ex1_comp1}
\end{subfigure}	
	\begin{subfigure}{.32\textwidth}
	\centering
	\includegraphics[width=5.5cm]{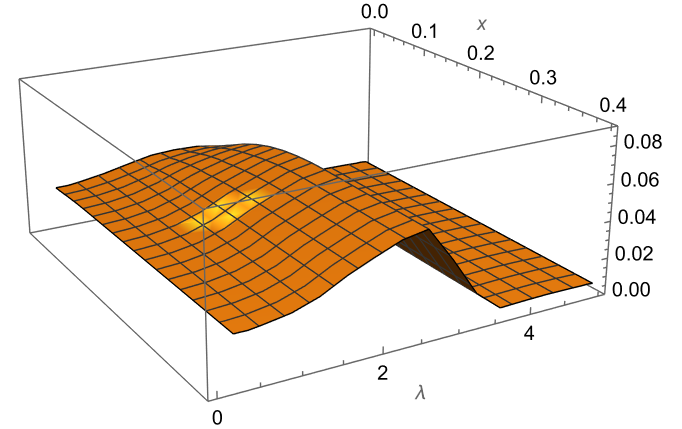}
	\caption{$V^2(x,\lambda)-V^{\text{NT}}(x,\lambda)$}
	\label{Ex1_tipp2}
\end{subfigure}
\begin{subfigure}{.32\textwidth}
	\centering
	\includegraphics[width=5.5cm]{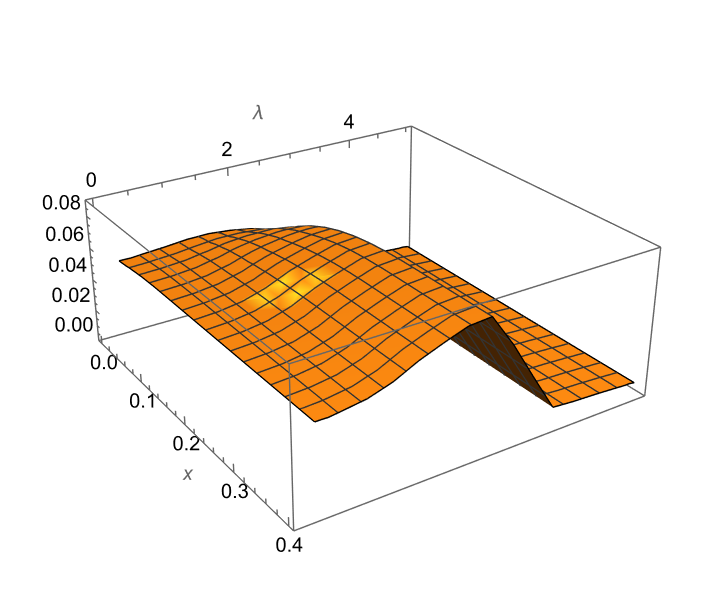}
	\caption{$V^1(x,\lambda)-V^{\text{NT}}(x,\lambda)$}
	\label{Ex1_tipp1}
\end{subfigure}
	\caption{{\small Comparison of the value function $V^2(x,\lambda_{\text{av}})$ (solid) and $V_{\text{CL}}(x,\lambda_{\text{av}})$ (dashed)  {as a function of $x$} (left), $V^2(x,\lambda)-V^{\text{NT}}(x,\lambda)$ (middle) and $V^1(x,\lambda)-V^{\text{NT}}(x,\lambda)$ ()right) in Example 1}}%
	\label{fig3}%
\end{figure}

Note that the premium rate is kept constant before the tipping point despite the changing value of momentary claim intensity $\lambda$ (so the insurance company is assuming and diversifying that risk over time), but it is raised at the moment of the tipping point to account for the average increase of the intensity at that point. Therefore, when we want to compare the attractiveness of a natural catastrophe element in the portfolio to the one of a 'regular' compound Poisson portfolio without catastrophe covers, a fair comparison seems to be the one where in the compound Poisson model we also raise the premium at the tipping point in the same fashion. In Figure \ref{Ex1_comp1}, 
$V^{2}(x,\lambda_{\text{av}})$ is compared with the corresponding value
function $V_{\text{CL}}(x,\lambda_{\text{av}})$ of the classical risk model analogue with the same constant intensity
$\lambda_{\text{av}}$ as the long-term average of the shot-noise process before and (with an increased value) also after the tipping point. Therefore the two receive the same premium amounts, both before and after the tipping point; for better visual comparison we start the shot-noise intensity in that same value $\lambda_0=\lambda_{\text{av}}$. {It was already shown in 
	\cite{AAM24} that the time variability of the claim intensity in the shot-noise scenario -- given that its value can be observed at every point in time and that the dynamics are known -- is in fact preferable for the shareholders in terms of the value function when compared to the situation with constant claim intensity (the safety loading for the policyholders staying at the same level). An interpretation is that this advantage comes from the additional degree of freedom of choosing the appropriate dividend payment strategy as a function of not only current surplus $x$, but also current intensity $\lambda$. Figure \ref{Ex1_comp1} shows that this stays true also when a tipping point is introduced.}
	Note that the dashed line in Figure \ref{Ex1_comp1} is not identical to the one in \cite[Fig.8.1(D)]{AAM24} (here the premium rate changes at the tipping point), although the two are very close. \\

Finally, we also look into the effects of the tipping point itself on a NatCat insurance portfolio. Figure \ref{Ex1_tipp2} plots the difference between the value function $V^2(x,\lambda)$ and the value function $V^{\text{NT}}(x,\lambda)$ of a process with the same initial dynamics, but without a tipping point coming later, across all relevant values of $x$ and $\lambda$. One sees that the difference is always positive. Intuitively, the additional variability of the claim intensity after the tipping point (due to the increased value of $\beta$) is an advantage, as was the shot-noise variation an advantage over a homogeneous Poisson portfolio. That is, from that point of view, the {presence} of a tipping point is an upward potential for shareholders, as long as the increased claim intensity {is} charged to the policyholders according to an expected value principle with the same risk loading. The difference is most pronounced for initial claim intensity values slightly above the long-term average $\lambda_{\text{av}}$ (which for this example is 1.20 before and 1.67 after the tipping point). Figure \ref{Ex1_tipp1} plots the difference  $V^1(x,\lambda)-V^{\text{NT}}(x,\lambda)$, which leads to larger (also always positive) values. The reason is that the value function $V^1$ is already larger, given that the transition to State 0 with its upward potential is already closer. One can also interpret Figures \ref{Ex1_2}, \ref{Ex1_val1} and \ref{Ex1_tipp1} as the situation of Example 1 with an Exp(1/3) instead of an Erlang (2,1/3) time horizon until the tipping point. \\

\subsection{Example 2: Erlang claims and increasing number of catastrophies}\label{sec7.2}

Consider a climate tipping point Erlang$\left(  2,1/3\right)  $ with the
frequency of catastrophes increasing from $\beta=1/6$ before the tipping point
to $\beta=1/5$ afterwards. Here, we consider an Erlang$\left(  2,1\right)  $
claim size distribution (i.e., $F_{U}(x)=1-(1+x)e^{-x}$, $x>0$) and {we switch to parameters} for which we know from Azcue
and Muler \cite{AM 2014} that a two-band strategy maximizes the
expected discounted dividend payments in the absence of shot-noise jumps in
the Poisson intensity. {Concretely}, we consider $\underline{\lambda}=10$, a
catastrophe shot-noise component with exponential intensity jumps with cdf
$F_{Y}(x)=1-e^{-x/2}$, $d=1/5$, a discount rate $q=1/10$ and $\eta=7/100$. We obtain
from (\ref{Definicion p}), that the premium rate equals $p=749/30$ before the
tipping point and the premium rate equals $p=642/25$ after the tipping point. So this example is the setting of Example 2 of \cite[Sec.8]{AAM24} after the tipping point, but we have a reduced shot-noise jump intensity (and correspondingly reduced premium rate) before the tipping point.\\

For the grid parameters, it suffices to consider a finite grid of
$60\times60$ points in the compact set $(x,\lambda)\in\left[  0,35\right]
\times\left[  10,30\right]  $ . Again, all points outside this compact set are defined to be in the action region. 

\begin{figure}[ptb]
	\begin{subfigure}{.32\textwidth}
		\centering
		\includegraphics[width=5.5cm]{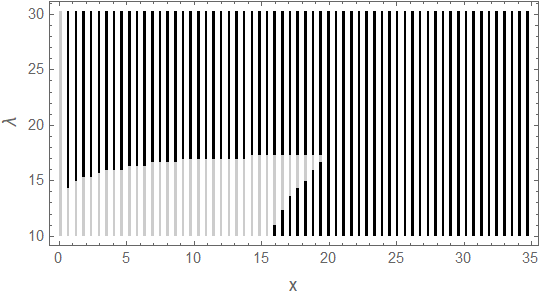}
		\caption{State 2}
		\label{Ex2_St2}
	\end{subfigure} \begin{subfigure}{.32\textwidth}
		\centering
		\includegraphics[width=5.5cm]{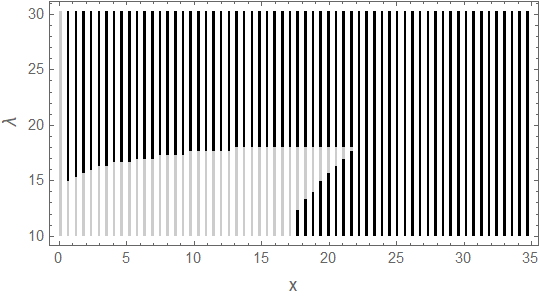}
		\caption{State 1}
		\label{Ex2_St1}
	\end{subfigure}
	\begin{subfigure}{.32\textwidth}
		\centering
		\includegraphics[width=5.5cm]{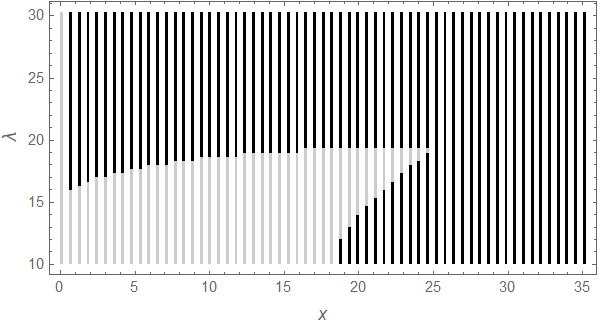}
		\caption{State 0}
		\label{Ex2_St0}
	\end{subfigure}
	\caption{{\small Action regions (dark) and no-action regions (light) of the optimal strategy as a function of $\lambda$ and $x$ for each of the states in Example 2}}%
	\label{fig4}%
\end{figure}

\begin{figure}[ptb]
	\begin{subfigure}{.32\textwidth}
		\centering
		\includegraphics[width=5.2cm]{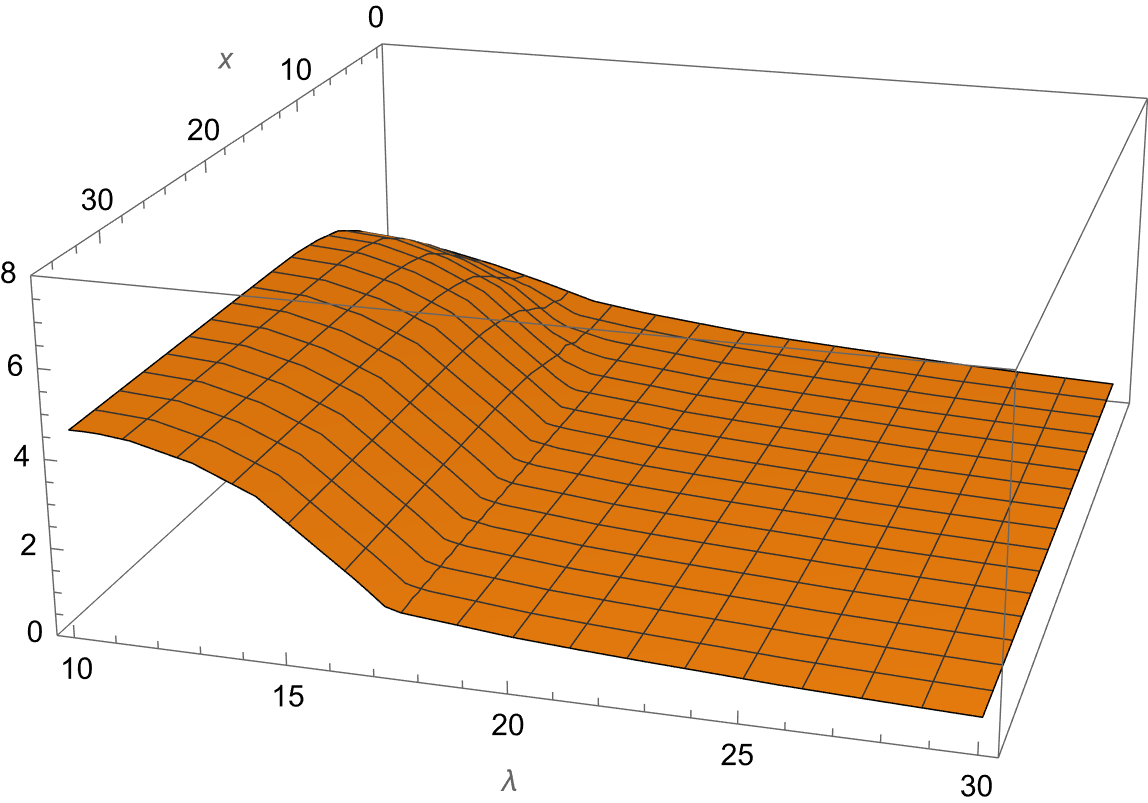}
		\caption{$V^2(x,\lambda)-x$}
	\end{subfigure} \begin{subfigure}{.32\textwidth}
		\centering
		\includegraphics[width=5.5cm]{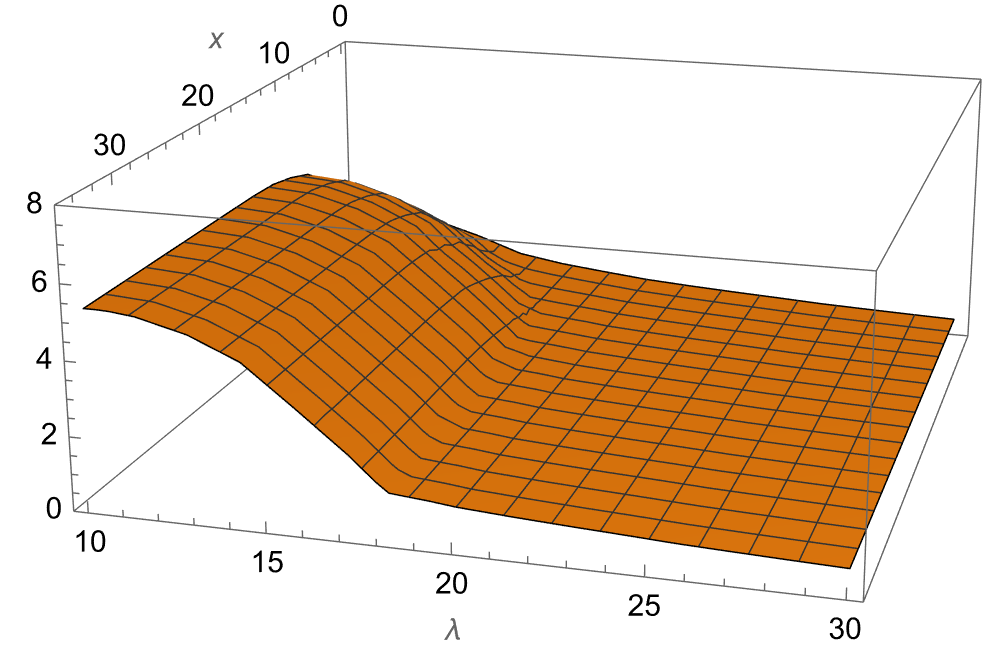}
		\caption{$V^1(x,\lambda)-x$}
	\end{subfigure}
	\begin{subfigure}{.32\textwidth}
		\centering
		\includegraphics[width=5.5cm]{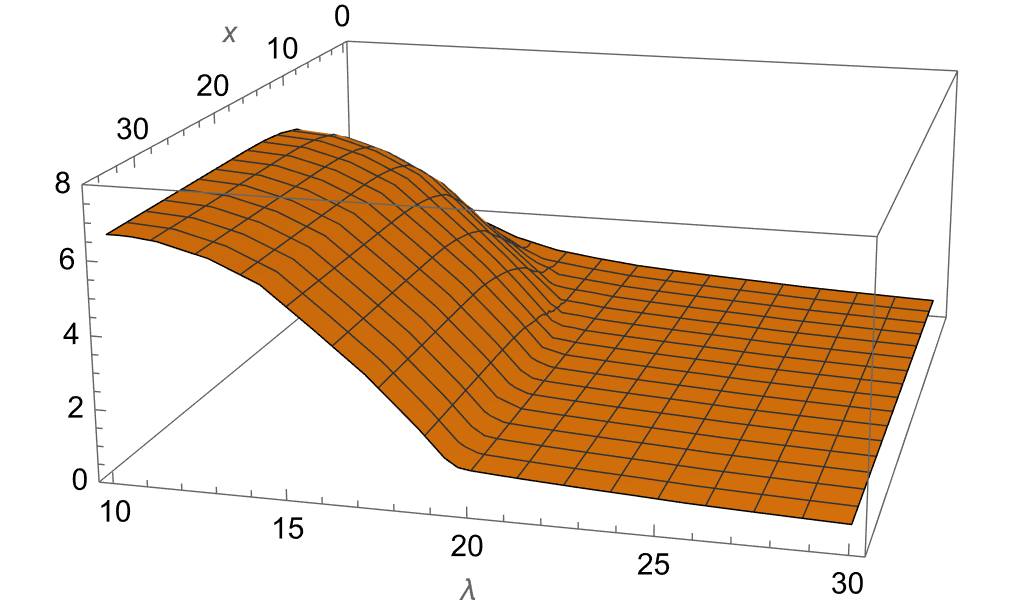}
		\caption{$V^0(x,\lambda)-x$}
	\end{subfigure}
	\caption{\small Value functions $V^i(x,\lambda)$ minus $x$ {as a function of $x$ and $\lambda$} for each State $i$ in Example 2}%
	\label{fig5}%
\end{figure}

Figure \ref{fig4} depicts the optimal action (dark) and no-action
(light) regions in the different states with respect to $x $ and
$\lambda$. Figure \ref{Ex2_St0} is identical to the respective regions plot of \cite[Fig.8.1(A)]{AAM24}, and one observes that also before the tipping point a similar pattern holds, with only mild adapations that are only of scaling type. Indeed, in each of the states  the
optimality of the two-band regime is retained for a small strip of values of $\lambda$. For smaller values of
$\lambda$ the two-band strategy collapses to a barrier
strategy, and for large values of $\lambda$ the optimal strategy is
to pay out all the surplus immediately as dividends. Figure \ref{fig5} depicts the respective optimal value function for each combination of $x$ and $\lambda$ in each of the three states {(plotted in excess of $x$)}. In this case the differences of the value function between the different states are relatively small. %

\begin{figure}[ptb]
	\centering
	\begin{subfigure}{.32\textwidth}
		\centering
		\includegraphics[width=5.5cm]{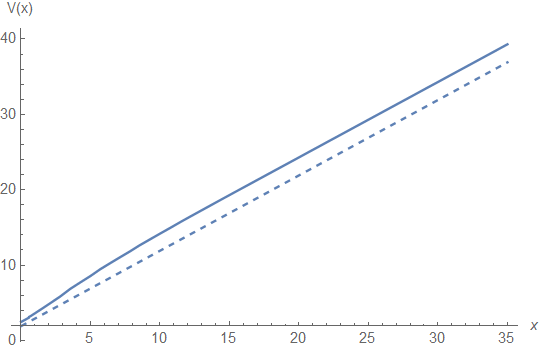}
		\caption{$V^2(x,\lambda_{\text{av}})$ vs. $V_{\text{CL}}(x,\lambda_{\text{av}})$}
		\label{Ex2_comp1}
	\end{subfigure}	
	\begin{subfigure}{.32\textwidth}
		\centering
		\includegraphics[width=5.5cm]{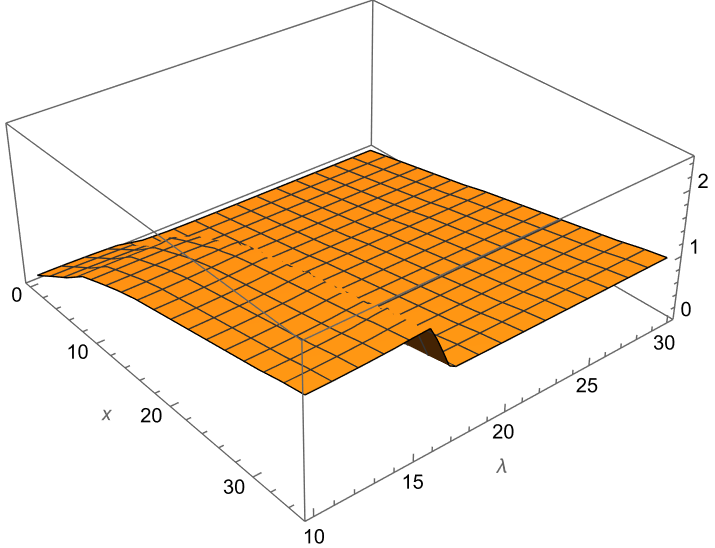}
		\caption{$V^2(x,\lambda)-V^{\text{NT}}(x,\lambda)$}
		\label{Ex2_tipp2}
	\end{subfigure}
		\begin{subfigure}{.32\textwidth}
		\centering
		\includegraphics[width=5.5cm]{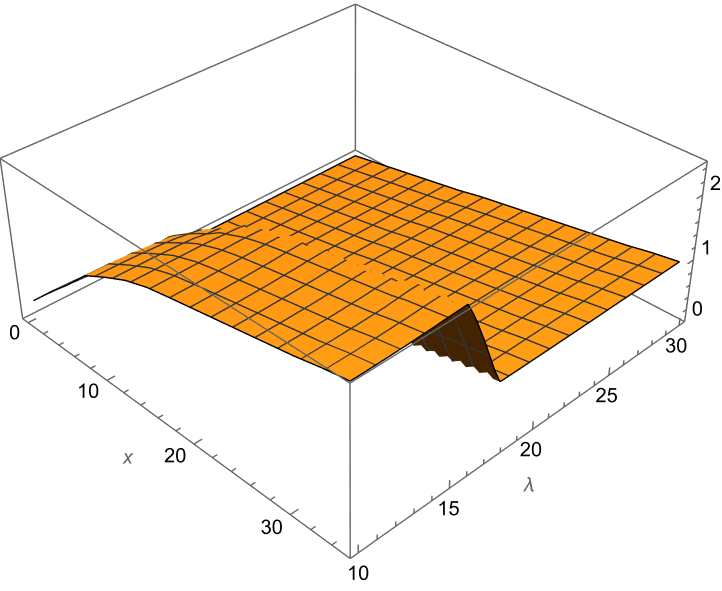}
		\caption{$V^1(x,\lambda)-V^{\text{NT}}(x,\lambda)$}
		\label{Ex2_tipp1}
	\end{subfigure}
	\caption{{\small Comparison of the value function $V^2(x,\lambda_{\text{av}})$ (solid) and $V_{\text{CL}}(x,\lambda_{\text{av}})$ (dashed)  {as a function of $x$} (left), $V^2(x,\lambda)-V^{\text{NT}}(x,\lambda)$ (middle) and $V^1(x,\lambda)-V^{\text{NT}}(x,\lambda)$ (right) in Example 2}}%
	\label{fig6}%
\end{figure}

Figure \ref{Ex2_comp1} compares $V^{2}(x,\lambda_{\text{av}})$ with the corresponding value
function $V_{\text{CL}}(x,\lambda_{\text{av}})$ of the respective classical risk model analogue with constant intensity
$\lambda_{\text{av}}$ and the same premium income. The
additional variability due to the shot-noise structure before and after the tipping point is again an advantage for the shareholders, although the difference is smaller in this case. In Figure \ref{Ex2_tipp2} we see that the presence of the tipping point itself in a compound shot-noise Cox process is also advantageous for the shareholders, and  Figure \ref{Ex2_tipp1} depicts the same situation for an Exp(1/3) time horizon. As in the previous example, this difference is larger than in Figure  \ref{Ex2_tipp2}.

\subsection{Example 3: Claim changes and increasing number of catastrophes}\label{sec7.3}
Next, let us consider a situation where before the climate
tipping time the claim sizes are exponential with the same parameters as in Example 1 and afterwards they are
deterministic at size 1/10. The latter corresponds to the situation of Example 3 in {\cite[Sec.8]{AAM24}} where no tipping point was considered. As  observed in that paper, the deterministic claim size assumption leads to an intricate action/no-action region pattern, and we are interested to see how the anticipation of the latter before the tipping point influences the shape of the optimal action regions. We assume here that the climate tipping point occurs at an exponentially distributed time Exp$(1/3)$, to avoid additional complication of an Erlang(2) time horizon (so here we only have two states). Assume $\underline{\lambda}=1/4$, an exponential intensity jump distribution with cdf $F_{Y}(x)=1-e^{-x/2}$, $d=7/10$, a discount rate $q=1/5$ and the
insurance safety loading applied in the policies equal to $\eta=1/5$. As in
Example 1, the tipping point increases the intensity of the intensity jump
 arrivals from $\beta=1/3$ before the tipping point to $\beta=1/2$
afterwards. 
The expected claim size before and after the tipping point remain the same. We then obtain from (\ref{Definicion p}), that the
premium rate equals $p=101/700$ before the tipping point and $p=141/700$ afterwards. For the grid parameters, the
finite grid of $80\times80$ points in the compact set $\left[  0,0.45\right]
\times\left[  1/4,5\right] $ is appropriate here (and all other points are defined to be in the
action region).

\begin{figure}[ptb]
\begin{center}
	\begin{subfigure}{.32\textwidth}
		\centering
		\includegraphics[width=5.5cm]{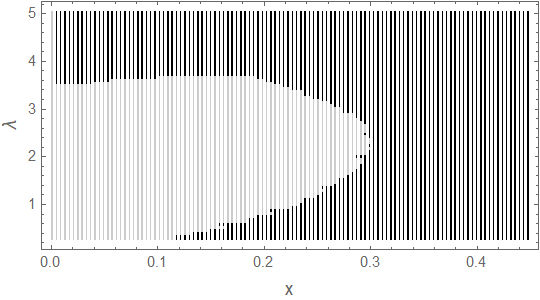}
		\caption{State 1}
		\label{Ex3_act1}
	\end{subfigure}\hspace{0.5cm
	}	\begin{subfigure}{.32\textwidth}
		\centering
		\includegraphics[width=5.5cm]{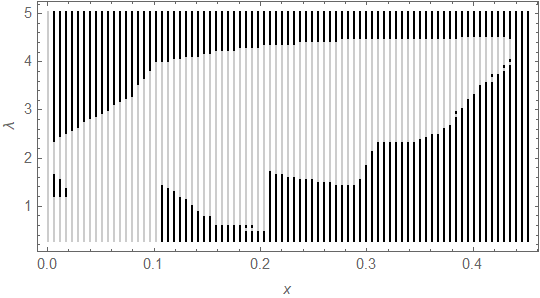}
		\caption{State 0}
		\label{Ex3_act0}
	\end{subfigure}
\end{center}
	\caption{{\small Action regions (dark) and no-action regions (light) of the optimal strategy as a function of $\lambda$ and $x$ for each of the states in Example 3}}%
	\label{fig7}%
\end{figure}

Figure \ref{fig7} depicts the action (dark) and no-action (light)
optimal regions in the different states with respect to $x$ and $\lambda$. We
see that the anticipation of a deterministic rather than exponential claim size distribution after the tipping point does not change the optimal strategy in the initial State 0 (Figures \ref{Ex3_act1} and \ref{Ex1_val1} coincide). Furthermore, Figure \ref{Ex3_act0} is identical with \cite[Fig.8.6(A)]{AAM24}. Figure \ref{Ex3_comp1} shows again that the additional variability of the intensity is an advantage for shareholders, and Figure \ref{Ex3_tipp1} depicts the advantage of the presence of a tipping point over the situation where the dynamics before the tipping point do not change in the future.  

\begin{figure}[ptb]
\begin{center}
	\begin{subfigure}{.32\textwidth}
		\centering
		\includegraphics[width=5.5cm]{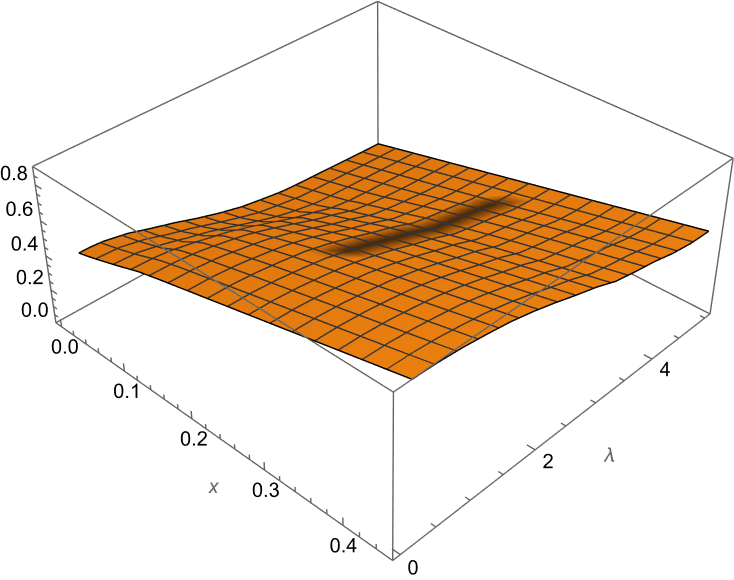}
		\caption{$V^1(x,\lambda)$}
		\label{Ex3_Val}
	\end{subfigure}\hspace{0.3cm}
	\begin{subfigure}{.32\textwidth}
		\centering
		\includegraphics[width=5.5cm]{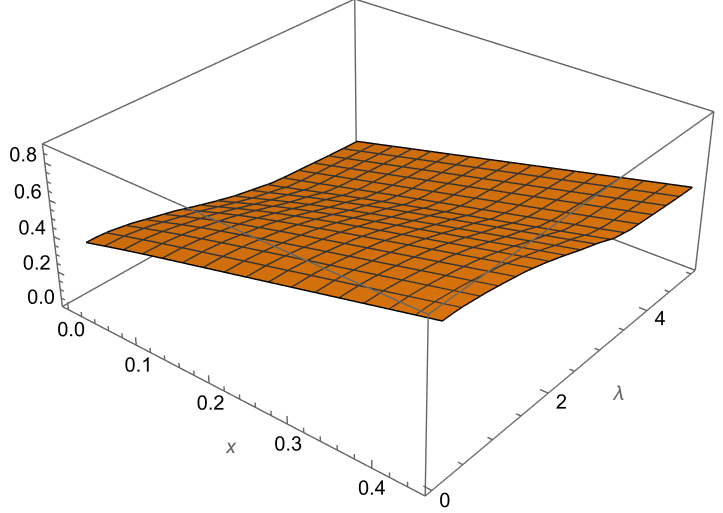}
		\caption{$V^0(x,\lambda)$}
		\label{Ex3_Comp}
	\end{subfigure}
\end{center}
	\caption{\small Value functions $V^i(x,\lambda)$ {as a function of $x$ and $\lambda$} for each State $i$ in Example 3}%
	\label{fig8}%
\end{figure}


\begin{figure}[ptb]
	\centering
	\begin{subfigure}{.32\textwidth}
		\centering
		\includegraphics[width=5.5cm]{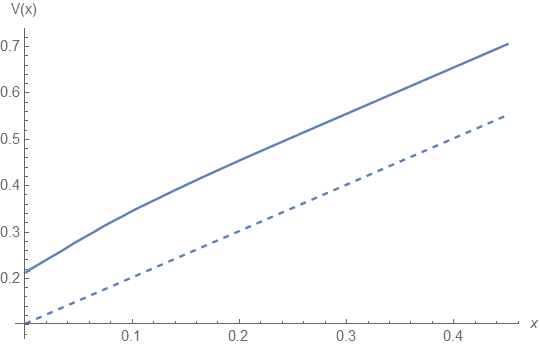}
		\caption{$V^1(x,\lambda_{\text{av}})$ vs. $V_{\text{CL}}(x,\lambda_{\text{av}})$}
		\label{Ex3_comp1}
	\end{subfigure}		\hspace{1cm}
	\begin{subfigure}{.32\textwidth}
		\centering
		\includegraphics[width=5.5cm]{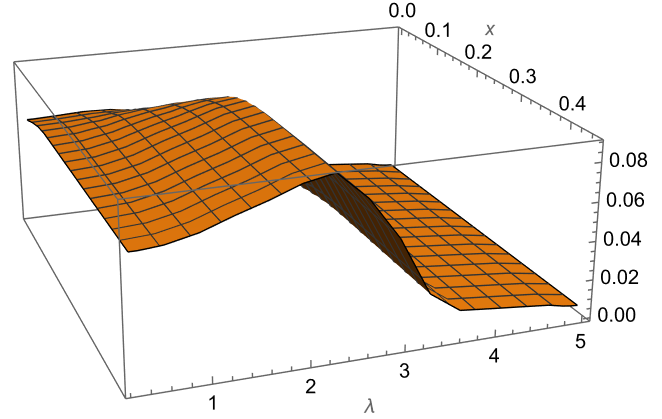}
		\caption{$V^1(x,\lambda)-V^{\text{NT}}(x,\lambda)$}
		\label{Ex3_tipp1}
	\end{subfigure}
	\caption{{\small Comparison of the value function $V^1(x,\lambda_{\text{av}})$ (solid) and $V_{\text{CL}}(x,\lambda_{\text{av}})$ (dashed)  {as a function of $x$} (left) and $V^2(x,\lambda)-V^{\text{NT}}(x,\lambda)$ in Example 3}}%
	\label{fig9}%
\end{figure}

%
%
%

\subsection{Example 4: A variant of Example 3 with truncated claims}\label{sec7.4}
One may question how the deterministic claims after the tipping point in Example 3 might be justified in practice. One such situation could be that the insurance policies contain in fact an upper coverage limit, and that after the tipping point the claims are all so large that the upper limit is in fact always reached (note that in that case, the uncertainty for which policyholders buy insurance (and pay a safety loading) is then solely the frequency and not the size of the claims). In the setting of Example 3, one would then have to assume that the claim size distribution before the tipping point is exponential, but truncated at some threshold, which in this example we choose to be 1/10. That is, $F_{U}(x)=\left(  1-e^{-10x}\right)
I_{x<1/10}+I_{x\geq1/10}$ before and $F_{U}(x)=I_{x\geq1/10}$ after the tipping point. All other assumptions and parameter values of Example 3 remain.  We then obtain from
(\ref{Definicion p}), that the premium rate equal $p=101(-1+e)/\left(
700 \,e\right) \approx 0.09 $ before the tipping point and $p=141/700\approx 0.20$ afterwards. For the grid parameters, the finite grid
of $80\times110$ points in the compact set $\left[  0,0.45\right]
\times\left[  1/4,7\right]  $ is appropriate here (with all remaining points being in the action
region).

\begin{figure}[ptb]
\begin{center}
		\begin{subfigure}{.32\textwidth}
		\centering
		\includegraphics[width=5.5cm]{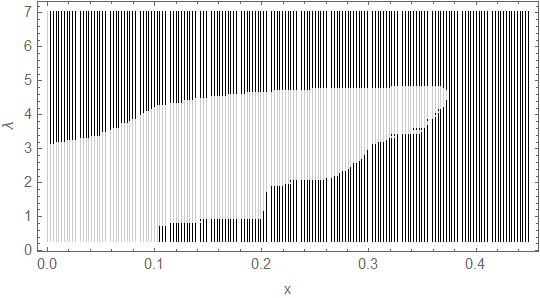}
		\caption{State 1}
		\label{Ex4_Str}
	\end{subfigure}\hspace{0.3cm} \begin{subfigure}{.32\textwidth}
		\centering
		\includegraphics[width=5.5cm]{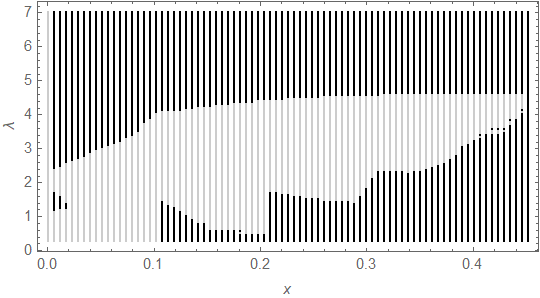}
		\caption{State 0}
		\label{Ex4_Val}
	\end{subfigure}
\end{center}
	\caption{{\small Action regions (dark) and no-action regions (light) of the optimal strategy as a function of $\lambda$ and $x$ for each of the states in Example 4}}%
	\label{fig10}%
\end{figure}

\begin{figure}[ptb]
\begin{center}
		\begin{subfigure}{.32\textwidth}
		\centering
		\includegraphics[width=5.5cm]{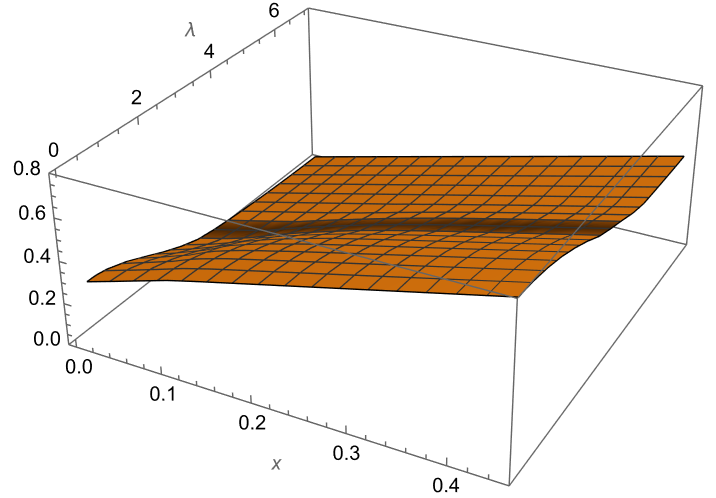}
		\caption{$V^1(x,\lambda)$}
		\label{Ex4_Str}
	\end{subfigure} \hspace{0.3cm}\begin{subfigure}{.32\textwidth}
		\centering
		\includegraphics[width=5.5cm]{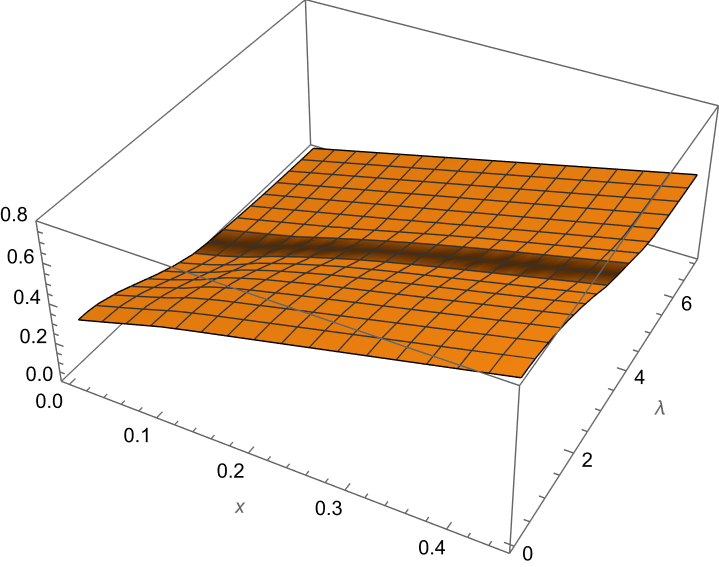}
		\caption{$V^0(x,\lambda)$}
		\label{Ex4_Val}
	\end{subfigure}
\end{center}
	\caption{\protect\small Value functions $V^i(x,\lambda)$ {as a function of $x$ and $\lambda$} for each State $i$ in Example 4}%
	\label{fig11}%
\end{figure}

\begin{figure}[ptb]
	\centering
	\begin{subfigure}{.32\textwidth}
		\centering
		\includegraphics[width=5.5cm]{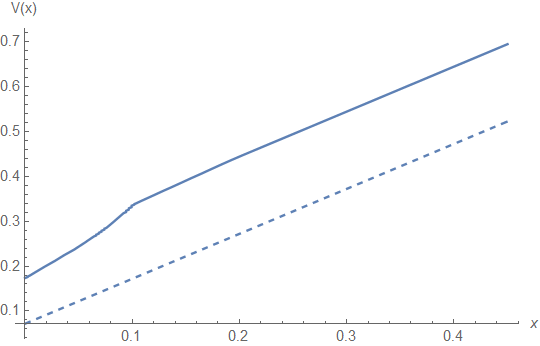}
		\caption{$V^1(x,\lambda_{\text{av}})$ vs. $V_{\text{CL}}(x,\lambda_{\text{av}})$}
		\label{Ex4_comp1}
	\end{subfigure}		\hspace{1cm}
	\begin{subfigure}{.32\textwidth}
		\centering
		\includegraphics[width=5.5cm]{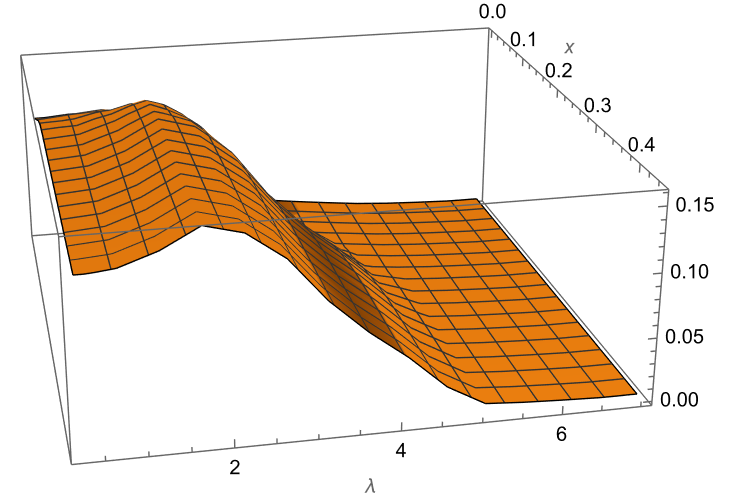}
		\caption{$V^1(x,\lambda)-V^{\text{NT}}(x,\lambda)$}
		\label{Ex4_tipp1}
	\end{subfigure}
	\caption{{\small Comparison of the value function $V^1(x,\lambda_{\text{av}})$ (solid) and $V_{\text{CL}}(x,\lambda_{\text{av}})$ (dashed)  {as a function of $x$} (left) and $V^2(x,\lambda)-V^{\text{NT}}(x,\lambda)$ in Example 4}}%
	\label{fig12}%
\end{figure}

Figure \ref{fig10} depicts the action and no-action optimal regions in the two states with respect to $x$ and $\lambda$, and Figure \ref{fig11} gives the respective value functions, which show a similar behavior as the ones of Example 3.  
Figure \ref{Ex4_comp1} compares the situation of the tipping point again with the one of a compound Poisson setup with constant intensity before and after the tipping point. Note here the kink at $x=1/10$, where (due to the deterministic claim amount of that size) the value function is continuous but not differentiable, so the notion of a viscosity solution is essential. Figure \ref{Ex4_tipp1} again assesses the quantitative advantage of having a tipping point for the shareholders for Example 4.  


\section{Conclusion}\label{concl}
In this paper we extended the analysis of \cite{AAM24} to the situation where an insurance company holding a NatCat portfolio faces a deterioration in the future, which we refer to as a tipping point. Under  {some} assumptions, we solved the resulting optimal control problem to identify the optimal dividend payment strategy that maximizes the expected aggregate discounted dividend payments {for shareholders} over the lifetime of the portfolio using viscosity solutions of the related Hamilton-Jacobi-Bellman equation. We proved uniform convergence of solutions on a finite grid of the parameter space and illustrate the optimal solutions and their properties for a number of concrete cases for which the number of catastrophes increases after the tipping point. It turns out that when the premiums are raised in a fair manner at that point (according to the same premium principle before and after the tipping point), the additional variability of the claim intensity can {turn out to be an advantage for} the shareholders. That is, under the assumption that all states and parameters are observable and that the model {accurately describes} reality, increased frequency and severity of catastrophes do not necessarily lead to more unattractive conditions for shareholders, and hence the insurance industry. It will be interesting to study in future research how certain model assumptions can be relaxed, and whether such a conclusion also holds in more general settings. 

\appendix
\section{Appendices}
\subsection{Proofs of Section \ref{Model and basic results}
	\label{Appendix Section Model and Basic Results}}

For technical reasons, let us define by $v^{\lambda}(x)$ the optimal value
function that maximizes the expected cumulative discounted dividend payouts
with constant-intensity $\lambda$ and premium rate $p$ (i.e., a
Cramer-Lundberg process) with switching value (at an exponential time
$\overline{\tau}\sim \text{Exp}(\xi)$ independent of the Cram\'er-Lundberg process)
$h^{\lambda}(x)=g(x,\lambda)$. It is straightforward to see the following
result, see for instance Wei, Yang, and Wang {\cite{WYW}} for a
similar problem. 

\begin{lemma}
	\label{Optima Unidimensional} $v^{\lambda}$ is non-decreasing with
	$v^{\lambda}(x)-x$ positive, non-decreasing and bounded. Moreover,
	$\lim_{\lambda\rightarrow\infty}v^{\lambda}(x)=x$ and $v^{\lambda}(x)\leq x+K$
	for some $K>0$
\end{lemma}

\textit{Proof of Proposition \ref{Basic Results V}:} We skip the proof that $V$ is non-increasing in $\lambda$ , uniformly
continuous in $\lambda$ and that $V(x,\lambda)\leq v^{\underline{\lambda}}(x)$ because the
proofs are rather cumbersome and similar to the ones of Proposition 3.1 in
{\cite{AAM24}}. Then, from Lemma \ref{Optima Unidimensional}, $V$
satisfies (\ref{Growth Condition}).

Let us show that $V(\cdot,\lambda):\left[  0,\infty\right)  \rightarrow
(0,\infty)$ is locally Lipschitz. It is straightforward to show that
$V(\cdot,\lambda)$ is non-decreasing because
\[
V(x_{2},\lambda)-V(x_{1},\lambda)\geq x_{2}-x_{1}.
\]
Let us prove the other inequality. Take $t_{0}=(x_{2}-x_{1})/p$ $<1$ and
consider $\lambda_{t_{0}}=\underline{\lambda}+e^{-dt_{0}}\left(
\lambda-\underline{\lambda}\right)  <\lambda$. Given an initial surplus
$x\geq0$ and $\varepsilon>0$, consider an admissible strategy $L=(L_{t}%
)_{t\geq0}\in\Pi_{x_{2},\lambda_{t_{0}}}$ such that
\[
J(L;x_{2},\lambda_{t_{0}})\geq V(x_{2},\lambda_{t_{0}})-\varepsilon
\]
for any $x_{2}>x_{1}$. Let us define the strategy $\widetilde{L}\in\Pi
_{x_{1},\lambda}$ that starting with surplus $x_{1}$, pays no dividends if
$X_{t}^{\widetilde{L}}<x_{2}$ and follows strategy $L$ after the current
reserve reaches $\left(  x_{2},\lambda_{t_{0}}\right)  $, hence%
\[
\widetilde{L}_{t}=\left\{
\begin{array}
	[c]{lll}%
	0 & \text{if} & t\leq t_{0}\text{ }\wedge\tau_{1}\wedge T_{1}\wedge
	\overline{\tau}\\
	L_{t-t_{0}} & \text{if} & t\geq t_{0}\text{ and }\tau_{1}\wedge T_{1}%
	\wedge\overline{\tau}>t_{0}\\
	0 & \text{if} & \tau_{1}\wedge T_{1}<t\ \text{and }\tau_{1}\wedge T_{1}\leq
	t_{0}\text{ }<\overline{\tau}.
\end{array}
\right.
\]
Therefore,
\[%
\begin{array}
	[c]{lll}%
	V(x_{1},\lambda) & \geq & J(\widetilde{L};x_{1},\lambda)\\
	& \geq & \mathbb{P}(\tau_{1}\wedge T_{1}\wedge\overline{\tau}>t_{0}%
	)J(L;x_{2},\lambda_{t_{0}})e^{-qt_{0}}\\
	& \geq & \left(  V(x_{2},\lambda_{t_{0}})-\varepsilon\right)  e^{-qt_{0}%
	}\mathbb{P}(\tau_{1}\wedge T_{1}\wedge\overline{\tau}>t_{0}),
\end{array}
\]
and%
\[%
\begin{array}
	[c]{ccc}%
	V(x_{2},\lambda)-V(x_{1},\lambda) & \leq & V(x_{2},\lambda)-\left(
	V(x_{2},\lambda_{t_{0}})-\varepsilon\right)  \mathbb{P}(\tau_{1}\wedge
	T_{1}\wedge\overline{\tau}>t_{0})e^{-qt_{0}}.%
\end{array}
\]
So, using that $V$ is non-increasing on $\lambda$ and $t_{0}<1,$%

\[%
\begin{array}
	[c]{lcl}%
	V(x_{2},\lambda)-V(x_{1},\lambda) & \leq & V(x_{2},\lambda)-V(x_{2}%
	,\lambda_{t_{0}})\mathbb{P}(\tau_{1}\wedge T_{1}\wedge\overline{\tau}%
	>t_{0})e^{-qt_{0}}\\
	& \leq & V(x_{2},\lambda)-V(x_{2},\lambda)\mathbb{P}(\tau_{1}\wedge
	T_{1}\wedge\overline{\tau}>t_{0})e^{-qt_{0}}\\
	& = & V(x_{2},\lambda)\left(  1-e^{-(q+\beta+\lambda+\xi)t_{0}}\right) \\
	& \leq & V(x_{2},\underline{\lambda})(q+\beta+\lambda+\xi)t_{0}\\
	& = & V(x_{2},\underline{\lambda})\frac{q+\beta+\lambda+\xi}{p}(x_{2}-x_{1})~
\end{array}
\]
which establishes the result.

Let us prove now that \ $V(x,\cdot)$ is locally Lipschitz in the open set
$(\underline{\lambda},\infty)$. Since $V$ is non-increasing on $\lambda$, we
have%
\[
0\leq V(x,\lambda_{1})-V(x,\lambda_{2}).
\]
Let us prove the other inequality. Take $L\in\Pi_{x,\lambda_{1}}$ such that
$V(x,\lambda_{1})\leq J(L;x,\lambda_{1})+\varepsilon$, and the associated
controlled process $(X_{t}^{L},\lambda_{t})$ starting at $(x,\lambda_{1}).$
Consider $\delta_{0}>0$ such that $\underline{\lambda}+e^{-d\delta_{0}}\left(
\lambda_{2}-\underline{\lambda}\right)  =\lambda_{1}$. Hence,%
\[
\delta_{0}=\frac{1}{d}\log(\frac{\lambda_{2}-\underline{\lambda}}{\lambda
	_{1}-\underline{\lambda}}).
\]
Since $\log(a)<a$ for $a>0,$ we also have that
\[
\delta_{0}\leq\frac{1}{d(\lambda_{1}-\underline{\lambda})}\left(  \lambda
_{2}-\lambda_{1}\right)  .
\]
Define $\widetilde{L}\in\Pi_{x,\lambda_{2}}$ as
\[
\widetilde{L}_{t}=\left\{
\begin{array}
	[c]{lll}%
	pt & \text{if} & t\leq\delta_{0}\text{ }\wedge\tau_{1}\wedge T_{1}%
	\wedge\overline{\tau}\\
	L_{t-\delta_{0}} & \text{if} & t\geq\delta_{0}\text{ and }\tau_{1}\wedge
	T_{1}\wedge\overline{\tau}>\delta_{0}\\
	p(\tau_{1}\wedge T_{1}) & \text{if} & t>\tau_{1}\wedge T_{1},t<\overline{\tau
	}\ \text{and }\delta_{0}\text{ }\geq\tau_{1}\wedge T_{1}.
\end{array}
\right.
\]
Since
\[
\int_{0}^{\delta_{0}}\text{ }\left(  \underline{\lambda}+e^{-ds}\left(
\lambda_{2}-\underline{\lambda}\right)  \right)  ds\leq\lambda_{2}\delta_{0},
\]
we obtain%

\[%
\begin{array}
	[c]{lll}%
	J(\widetilde{L};x,\lambda_{2}) & \geq & J(L;x,\lambda_{1})e^{-q\delta_{0}%
	}\mathbb{P}(\tau_{1}\wedge T_{1}\wedge\overline{\tau}>\delta_{0})\\
	& \geq & J(L;x,\lambda_{1})e^{-(q+\beta+\lambda+\xi)\delta_{0}}\\
	& \geq & J(L;x,\lambda_{1})\left(  1-(q+\beta+\lambda+\xi)\delta_{0}\right)  .
\end{array}
\]
As a result, taking $\lambda_{2}\ $and$~\lambda_{1}$ close enough so that 
$\delta_{0}<1$,
\[%
\begin{array}
	[c]{lll}%
	V(x,\lambda_{1})-V(x,\lambda_{2}) & \leq & J(L;x,\lambda_{1})-J(\widetilde{L}%
	,x,\lambda_{2})+\varepsilon\\
	& \leq & J(L;x,\lambda_{1})(q+\beta+\lambda+\xi)\delta_{0}+\varepsilon\\
	& \leq & V(x,\lambda_{1})\allowbreak\frac{q+\beta+\lambda+\xi}{d(\lambda
		_{1}-\underline{\lambda})}\left(  \lambda_{2}-\lambda_{1}\right)
	+\varepsilon,
\end{array}
\]
and the result follows. At the lower boundary $\underline{\lambda}$ we only have
the uniform continuity result because the Lipschitz constant blows up as
$\lambda\searrow\underline{\lambda}$.\hfill $\blacksquare$\\

\textit{Proof of Proposition \ref{Convergencia puntual con lambda a infinito}:} For 
$\lambda$ large enough, consider $t\left(  \lambda\right)  $ so $\underline{\lambda}+e^{-dt\left(
	\lambda\right)  }\left(  \lambda-\underline{\lambda}\right)  =\ln(\lambda)$. Hence,%

\[
t\left(  \lambda\right)  :=\frac{1}{d}\ln(\frac{\lambda-\underline{\lambda}%
}{\ln(\lambda)-\underline{\lambda}}).
\]
Take a near optimal strategy $L=(L_{t})_{t\geq0}\in\Pi_{x,\lambda}$ such that
$V(x,\lambda)\leq J(L;x,\lambda)+\varepsilon$, by definition $\lambda_{t}%
\geq\lambda_{t}^{c}=$ $\underline{\lambda}+e^{-dt}\left(  \lambda
-\underline{\lambda}\right)  $, and so $\lambda_{t}\geq\ln(\lambda)$ for
$t\leq t\left(  \lambda\right)  $. Then, using Lemma \ref{DPP}, Lemma
\ref{Optima Unidimensional} and Proposition \ref{Basic Results V},%

\[%
\begin{array}
	[c]{lll}%
	V(x,\lambda)-\varepsilon & \leq & J(L;x,\lambda)\\
	& = &
	\begin{array}
		[c]{l}%
		\mathbb{E}(%
		{\textstyle\int\nolimits_{0^{-}}^{t\left(  \lambda\right)  \wedge
				\overline{\tau}\wedge\tau^{L}}}
		e^{-qs}dL_{s}+I_{\{\overline{\tau}\wedge t^{{}}\left(  \lambda\right)
			\wedge\tau^{L}=\overline{\tau}\}}g(X_{\overline{\tau}}^{L},\lambda
		_{\overline{\tau}}))\\
		+e^{-qt\left(  \lambda\right)  }\mathbb{E}\left(  I_{\{\overline{\tau}%
			\wedge\tau\wedge\tau^{L}=t\left(  \lambda\right)  \}}V(X_{t\left(
			\lambda\right)  }^{L},\lambda_{t\left(  \lambda\right)  })\right)
	\end{array}
	\\
	& \leq & v^{\ln(\lambda)}(x)+e^{-qt\left(  \lambda\right)  }%
	v^{\underline{\lambda}}(x+pt\left(  \lambda\right)  ).
\end{array}
\]
Since $t\left(  \lambda\right)  \rightarrow\infty$ as $\lambda\rightarrow
\infty$ and by Lemma \ref{Optima Unidimensional}, we have that $\lim
_{\lambda\rightarrow\infty}v^{\ln(\lambda)}(x)=x$, we obtain for some $K>0,$
\[
\lim_{\lambda\rightarrow\infty}e^{-qt\left(  \lambda\right)  }%
v^{\underline{\lambda}}(x+pt\left(  \lambda\right)  )\leq\lim_{\lambda
	\rightarrow\infty}e^{-qt\left(  \lambda\right)  }\left(  x+pt\left(
\lambda\right)  +K\right)  =0,
\]
hence we have the result. \hfill $\blacksquare$

\subsection{Proofs of Section \ref{Seccion Viscosidad}~
	\label{Appendix Section HJB}}

\textit{Proof of Proposition \ref{Prop V is a viscosity solution}:} Let us show that $V$ is a viscosity supersolution. Given initial values
$\left(  x,\lambda\right)  \in(0,\infty)\times(\underline{\lambda},\infty) $
and any $l$ $\geq0$, we consider the admissible strategy $L\in\Pi_{x,\lambda}$
where the company pays dividends with constant rate $l$ until $\tau^{L}$
$\wedge\overline{\tau}$. Let $\varphi$ be a test function for supersolution of
(\ref{HJB}) at $(x,\lambda)$, the associated controlled surplus is given by
$X_{t}^{L}=x+\left(  p-l\right)  t$ and the intensity process is $\lambda
_{t}=\underline{\lambda}+e^{-dt}\left(  \lambda-\underline{\lambda}\right)  $
for $t<\tau_{1}\wedge T_{1}\wedge\overline{\tau}$. Applying Lemma \ref{DPP}
with stopping time $\tau_{1}\wedge T_{1}\wedge\overline{\tau}\wedge h$, we
have%
\[%
\begin{array}
	[c]{lll}%
	0 & = & V(x,\lambda)-\varphi(x,\lambda)\\
	& \geq & \mathbb{E}(\int_{0}^{\tau_{1}\wedge T_{1}\wedge\overline{\tau}\wedge
		h}e^{-qt}ldt)\\
	&  & +\mathbb{E}\left(  I_{\left\{  \tau_{1}<T_{1}\wedge\overline{\tau}\wedge
		h\right\}  }e^{-q\tau_{1}}\varphi(x+(p-l)\tau_{1}-U_{1},\underline{\lambda
	}+e^{-d\tau_{1}}\left(  \lambda-\underline{\lambda}\right)  )\right) \\
	&  & +\mathbb{E}\left(  I_{\left\{  T_{1}<\tau_{1}\wedge\overline{\tau}\wedge
		h\right\}  }e^{-qT_{1}}\varphi(x+(p-l)T_{1},\underline{\lambda}+e^{-dT_{1}%
	}\left(  \lambda-\underline{\lambda}\right)  +Y_{1}))\right) \\
	&  & +\mathbb{E}\left(  I_{\left\{  h<\tau_{1}\wedge\overline{\tau}\wedge
		T_{1}\right\}  }e^{-qh}\varphi(x+(p-l)h,\underline{\lambda}+e^{-dh}\left(
	\lambda-\underline{\lambda}\right)  )\right) \\
	&  & +\mathbb{E}\left(  I_{\left\{  \overline{\tau}<\tau_{1}\wedge T_{1}\wedge
		h\right\}  }e^{-q\overline{\tau}}g(x+(p-l)\overline{\tau},\underline{\lambda
	}+e^{-d\overline{\tau}}\left(  \lambda-\underline{\lambda}\right)  )\right)
	-\varphi(x,\lambda).
\end{array}
\]
So, dividing by $h$ and taking $h\rightarrow0^{+}$, and taking into account
that $\tau_{1},T_{1}$ and $\overline{\tau}$ are independent exponential  random
variables, we obtain%
\[
\mathcal{L}(\varphi)(x,\lambda)+l(1-\varphi_{x}(x,\lambda))\leq0.
\]
Hence, taking $l\rightarrow\infty,$we get%
\[
\max\{\mathcal{L}(\mathbb{\varphi})(x,\lambda),1-\mathbb{\varphi}%
_{x}(x,\lambda)\}\leq0.
\]
This proves that $V$ is a viscosity supersolution at $(x,\lambda)$.

We omit the proof that $V$ is a viscosity subsolution. This result is similar
to the one in Proposition 4.1 in {\cite{AAM24}}.\hfill $\blacksquare$\\

\noindent The next lemma will be used to prove Proposition \ref{MenorSuper}, and the proof is similar to the one of Lemma A.3 in {\cite{AAM24}} in
which $\xi=0$.

\begin{lemma}
	\label{A.1} Let $\overline{u}$\ be a non-negative supersolution of (\ref{HJB})
	satisfying the growth condition (\ref{Growth Condition}) and non-increasing on
	$\lambda$.\ Then, the sequence of positive convolution functions $\overline
	{u}^{m}:(0,\infty)\times(\underline{\lambda},\infty)\rightarrow\mathbb{R}%
	$\ defined as
	\[
	\overline{u}^{m}(x,\lambda)=m^{2}\int\nolimits_{-\infty}^{\infty}%
	\int\nolimits_{-\infty}^{\infty}\overline{u}(x+s,\lambda+t)\phi(ms)\phi
	(mt)\,ds\,dt,
	\]
	where $\phi$ is a non-negative continuously differentiable function with
	support included in $(0,1)$ such that $\int_{0}^{1}\phi(x)dx=1$, satisfies
	
	(a) $\overline{u}^{m}$\ is continuously differentiable.
	
	(b) $\overline{u}^{m}(x,\lambda)\leq K+x$ and $\overline{u}^{m}$ is
	non-increasing on $\lambda$.
	
	(c)$\ 1\leq\overline{u}_{x}^{m}\leq(\left(  q+\lambda+\beta+\xi\right)
	/p)\overline{u}(x+1/m,\lambda)$ in $(0,\infty)\times(\underline{\lambda
	},\infty)$.
	
	(d) $\overline{u}^{m}$\ $\rightarrow\overline{u}$\ uniformly on compact sets
	in $(0,\infty)\times(\underline{\lambda},\infty)$ and $\nabla\overline{u}^{m}
	$\ converges to $\nabla\overline{u}$\ a.e. in $(0,\infty)\times
	(\underline{\lambda},\infty)$.
	
	(e) There exists a sequence $c_{m}$ with $\lim\limits_{m\rightarrow\infty
	}c_{m}=0$ such that
	\[
	\sup\nolimits_{\left(  x,\lambda\right)  \in A_{0}}\mathcal{L}(\overline
	{u}^{m})\left(  x,\lambda\right)  \leq c_{m},\text{where }A_{0}=[0,x_{0}%
	]\times\lbrack\lambda_{0},\lambda_{1}],
	\]
	where $x_{0}>0$ and $\lambda_{0},\lambda_{1}\in(\underline{\lambda},\infty).$
\end{lemma}

\textit{Proof of Proposition \ref{MenorSuper}:} {Consider} $\overline{u}$ a non-negative supersolution of (\ref{HJB})
satisfying the growth condition (\ref{Growth Condition}). Take any admissible
strategy $L\in\Pi_{x,\lambda}$, and $\left(  X_{t}^{L},\lambda_{t}\right)  $
as the corresponding controlled risk process starting at $\left(
x,\lambda\right)  $. Take $\overline{u}^{m}$ as defined in Lemma \ref{A.1} in
$(0,\infty)\times(\underline{\lambda},\infty)$ and let us extend this function
as $\overline{u}^{m}(x,\lambda)=0.$ As in the proof of Proposition
4.2\textit{\ }in {\cite{AAM24}}, but considering the switching
time\textit{\ }$\overline{\tau}$, we have (using $\overline{u}_{x}^{m}\geq1)$,%

\begin{equation}%
	\begin{array}
		[c]{l}%
		\mathbb{E}\left(  I_{t\wedge\tau^{L}<\overline{\tau}}\overline{u}%
		^{m}(X_{t\wedge\tau^{L}}^{L},\lambda_{t\wedge\tau^{L}})e^{-q(t\wedge\tau^{L}%
			)}+I_{\overline{\tau}<t\wedge\tau^{L}}e^{-q\overline{\tau}}g(X_{\overline
			{\tau}},\lambda_{\overline{\tau}})\right)  -\overline{u}^{m}(x,\lambda)\\
		\leq\mathbb{E}\left(  \int\nolimits_{0}^{\overline{\tau}\wedge t\wedge\tau
			^{L}}\mathcal{L}(\overline{u}^{m})(X_{s^{-}}^{L},\lambda_{s^{-}}%
		)e^{-qs}ds\right)  -\mathbb{E}\left(  \int\nolimits_{0^ -}^{\overline{\tau}\wedge
			t\wedge\tau^{L}}e^{-qs}dL_{s}\right)  \text{.}%
	\end{array}
	\label{ItoUnMenorSuper}%
\end{equation}
By the monotone convergence theorem and since $L_{t}$ is a non-decreasing
process, we obtain
\begin{equation}
	\lim\limits_{t\rightarrow\infty}\mathbb{E}\left(  \int\nolimits_{0^{-}%
	}^{\overline{\tau}\wedge t\wedge\tau^{L}}e^{-qs}dL_{s}\right)  =\mathbb{E}%
	\left(  \int\nolimits_{0^{-}}^{\overline{\tau}\wedge\tau^{L}}e^{-qs}%
	dL_{s}\right)  .\nonumber
\end{equation}
We get from Lemma \ref{A.1}-(c)%
\begin{equation}
	-(q+\lambda+\beta+\xi)\overline{u}(x+\frac{1}{m},\lambda)\leq\mathcal{L}%
	(\overline{u}^{m})(x,\lambda)\leq\left(  q+\lambda+\beta+\xi\right)
	\overline{u}(x+\frac{1}{m},\lambda)+\beta\overline{u}(x,\lambda)+{\small \xi
		g(x,\lambda)}.\label{acotacion L}%
\end{equation}
From Lemma \ref{A.1}-(b), the growth condition (\ref{Growth Condition}) on
$\overline{u}$ and $g$, and since $X_{s}^{L}\leq x+ps$, $\lambda
_{s}>\underline{\lambda}$ we obtain
\begin{equation}
	\overline{u}(X_{s}^{L},\lambda_{s})\leq K+x+ps\text{, }g(X_{s},\lambda
	_{s})\leq K+x+ps\label{Acotacionu}%
\end{equation}
for some $K>0$. Therefore, by the bounded convergence theorem, we have
\begin{align}
	&  \lim\limits_{t\rightarrow\infty}\mathbb{E}\left(  \int\nolimits_{0}%
	^{\overline{\tau}\wedge t\wedge\tau^{L}}\mathcal{L}(\overline{u}^{m}%
	)(X_{s^{-}}^{L},\lambda_{s^{-}})e^{-qs}ds+I_{\overline{\tau}\wedge t<\tau^{L}%
	}\left(  e^{-q\overline{\tau}}g(X_{\overline{\tau}},\lambda_{\overline{\tau}%
	})\right)  \right) \label{monotone2}\\
	&  =\mathbb{E}\left(  \int\nolimits_{0}^{\overline{\tau}\wedge\tau^{L}%
	}\mathcal{L}(\overline{u}^{m})(X_{s^{-}}^{L},\lambda_{s^{-}})e^{-qs}%
	ds+I_{\overline{\tau}<\tau^{L}}\left(  e^{-q\overline{\tau}}g(X_{\overline
		{\tau}},\lambda_{\overline{\tau}})\right)  \right)  .\nonumber
\end{align}
From (\ref{ItoUnMenorSuper}) and (\ref{monotone2}), we get%

\begin{equation}%
	\begin{array}
		[c]{l}%
		\lim\limits_{t\rightarrow\infty}\mathbb{E}\left(  I_{t\wedge\tau^{L}%
			<\overline{\tau}}\overline{u}^{m}(X_{t\wedge\tau^{L}}^{L},\lambda_{t\wedge
			\tau^{L}})e^{-q(t\wedge\tau^{L})}\right)  -\overline{u}^{m}(x,\lambda)\\
		=-\overline{u}^{m}(x,\lambda)\\
		\leq\mathbb{E}\left(  \int\nolimits_{0}^{\tau^{L}}\mathcal{L}(\overline{u}%
		^{m})(X_{s^{-}}^{L},\lambda_{s^{-}})e^{-qs}ds\right)  -\mathbb{E}\left(
		\int\nolimits_{0^-}^{\overline{\tau}\wedge\tau^{L}}e^{-qs}dL_{s}+I_{\overline
			{\tau}<\tau^{L}}e^{-q\overline{\tau}}g(X_{\overline{\tau}},\lambda
		_{\overline{\tau}})\right) \\
		=\mathbb{E}\left(  \int\nolimits_{0}^{\tau^{L}}\mathcal{L}(\overline{u}%
		^{m})(X_{s^{-}}^{L},\lambda_{s^{-}})e^{-qs}ds\right)  -J(L;x,\lambda).
	\end{array}
	\label{limite0}%
\end{equation}
We now show that
\begin{equation}
	\limsup\limits_{m\rightarrow\infty}\mathbb{E}\left(  \int\nolimits_{0}%
	^{\tau^{L}}\mathcal{L}(\overline{u}^{m})(X_{s^{-}}^{L},\lambda_{s^{-}}%
	)e^{-qs}ds\right)  \leq0.\label{limite2}%
\end{equation}
Since $\overline{u}^{m}(x,\lambda)\leq K+x$, Lemma \ref{A.1}-(b) and (c),%

\[%
\begin{array}
	[c]{lll}%
	\left\vert \mathcal{L}(\overline{u}^{m})(X_{s^{-}}^{L},\lambda_{s^{-}%
	})\right\vert  & \leq & \left(  q+\lambda+\beta+\xi\right)  \overline
	{u}(X_{s^{-}}^{L}+\frac{1}{m},\lambda_{s^{-}})+\beta\overline{u}(X_{s^{-}}%
	^{L},\lambda_{s^{-}})\\
	& \leq & \left(  q+\lambda+\beta+\xi\right)  \overline{u}(X_{s^{-}}^{L}%
	+\frac{1}{m},\underline{\lambda})+\beta\overline{u}(X_{s^{-}}^{L}%
	,\underline{\lambda})\\
	& \leq & \left(  q+\lambda+2\beta+\xi\right)  \left(  K+x+1+ps\right)  .
\end{array}
\]
As a consequence, given any $\varepsilon>0,$ there exists $T$ large enough
such that
\begin{equation}
	\int\nolimits_{T}^{\infty}\mathcal{L}(\overline{u}^{m})(X_{s^{-}}^{L}%
	,\lambda_{s^{-}})e^{-qs}ds<\frac{\varepsilon}{4}\label{arreglado5}%
\end{equation}
for any $m\geq1.$ For $s\leq T$, the controlled process $X_{s^{-}}^{L}$ $\leq
x_{0}:=x+pT$ , $\lambda_{s}$ $\geq\lambda_{T}:=\underline{\lambda}+\left(
\lambda-\underline{\lambda}\right)  e^{-dT}$. \ Since $\boldsymbol{\lambda}$
is a piecewise decreasing process that has upward jumps at exponential times
with jumps of finite mean, it is possible to find a constant
$l$\ large enough such that
\[
\mathbb{P}(\sup_{s\leq T}\lambda_{s}>l)<\frac{\varepsilon}{4\left(
	q+\lambda+2\beta+\xi\right)  \left(  x+\frac{1}{m}+ps\right)  }\,.
\]
Now, take the compact set $\left[  0,x_{0}\right]  \times\left[  \lambda
_{T},l\right]  $ and the set%

\[
C=\left\{  \left(  \lambda_{s}\right)  _{s\leq T}:\sup_{s\leq T}\lambda
_{s}\leq M\right\}  .
\]
From Lemma \ref{A.1}-(e), there exists $m_{0}$ large enough such that for any
$m\geq m_{0},$%
\[
I_{C}\int\nolimits_{0}^{T}\mathcal{L}(\overline{u}^{m})(X_{s^{-}}^{L}%
,\lambda_{s^{-}})e^{-qs}ds\leq c_{m}\int\nolimits_{0}^{T}e^{-qs}ds\leq
\frac{c_{m}}{q}\leq\frac{\varepsilon}{2},
\]
and so taking the expectation we get (\ref{limite2}). Then, from
(\ref{limite0}) and using Lemma \ref{A.1}-(d), we obtain%

\begin{equation}
	\overline{u}(x,\lambda)=\lim\limits_{m\rightarrow\infty}\overline{u}%
	^{m}(x,\lambda)\geq J(L;x,\lambda).\label{supersolucionMayorqueVL}%
\end{equation}
As a consequence, since $V$ is a viscosity solution of (\ref{HJB}), the result
follows. \hfill $\blacksquare$

\subsection{Proofs of Section \ref{Section Asymptotic} \label{Appendix Asymptotic}%
}

\textit{Proof of Proposition \ref{Proposicion acotacion de retirar}:} Let us define for $M\geq\widehat{x}>0$, where $\widehat{x}$ is defined in
Assumption (A1),
\[
V_{M}(x,\lambda)=\left\{
\begin{array}
	[c]{lll}%
	V(x,\lambda) & \text{if} & x\leq M,\\
	V(M,\lambda)+x-M & \text{if} & x>M.
\end{array}
\right.
\]
Note that $V_{M}(x,\lambda)$ is a limit of value functions of admissible
strategies in $\Pi_{x,\lambda}$. $V_{M}$ is locally Lipschitz, increasing on
$x$ and non-increasing on $\lambda$ (see Proposition \ref{Basic Results V}).
Also, $V_{M}$ is a viscosity supersolution of (\ref{HJB}) for $x<M$. \ In
addition, $\partial_{x}V_{M}(x,\lambda)=1$ for $x>M$. Let us show that $V_{M}$
is a viscosity supersolution of (\ref{HJB}) for $x>M$ taking $M$ large enough.
It is enough to show that $\mathcal{L}(V_{M})(x,\lambda)\leq0 $ for $x>M$. We have%

\[%
\begin{array}
	[c]{l}%
	\mathcal{L}(V_{M})(x,\lambda)\\
	\leq p-(q+\lambda)V_{M}(x,\lambda)+\lambda%
	{\textstyle\int\nolimits_{0}^{x}}
	V_{M}(x-\alpha,\lambda)dF_{U}(\alpha)+{\small \xi}\left(
	{\small g(\widehat{x},\lambda)+M-\widehat{x}-V(M,\lambda)}\right) \\
	\leq p-(q+\lambda)(V(M,\lambda)+x-M)+\lambda%
	{\textstyle\int\nolimits_{0}^{x}}
	(V(M,\lambda)+x-\alpha-M)dF_{U}(\alpha)+{\small \xi}\left(
	{\small g(\widehat{x},\lambda)+M-\widehat{x}-V(M,\lambda)}\right) \\
	\leq p-qV(M,\lambda)+{\small \xi}\left(  {\small g(\widehat{x},\lambda
		)+M-\widehat{x}-V(M,\lambda)}\right) \\
	\leq p-qV(M,\lambda)+{\small \xi}\left(  {\small g(\widehat{x},\lambda
		)-}\widehat{x}\right).
\end{array}
\]
And so, since $V(x,\lambda)\geq x$,%
\[
\mathcal{L}(V_{M})(x,\lambda)\leq p-qM+{\small \xi}\left(  {\small g(}%
\widehat{x}{\small ,\lambda)-}\widehat{x}\right)  .
\]
Hence, taking $M\geq\overline{x}$, we obtain $\mathcal{L}(V_{M})(x,\lambda
)\leq0$ for $x>M$.

\noindent Let us define%

\[
B:=\{(x,\lambda)\in\left[  0,\infty\right)  \times\left[  \underline{\lambda
},\infty\right)  :V~\text{is differentiable}\}.
\]
$B$ is a dense set in $\left[  0,\infty\right)  \times\left[
\underline{\lambda},\infty\right)  $ and for any point $(M,\lambda)\in B$ with
$M\geq\overline{x}$ we have that $\partial_{x}V_{M}(M,\lambda)=1$. Consequently, by
Corollary \ref{VerificacionDividendos}, we obtain that $V_{M}$ coincides with
$V.$\hfill $\blacksquare$\\

\textit{Proof of Proposition \ref{Convergencia Uniforme con Lambda}:} We have that $V(x,\lambda)=x+a(\lambda)$ for $x\geq\overline{x}$ as a
consequence of Proposition \ref{Proposicion acotacion de retirar}, taking
\[
a(\lambda):=V(\overline{x},\lambda)-\overline{x}\text{. }%
\]
Also, from Proposition \ref{Convergencia puntual con lambda a infinito},
$\lim_{\lambda\rightarrow\infty}V(\overline{x},\lambda)-\overline{x}=0$ and
$V(\overline{x},\lambda)$ is non-increasing on $\lambda$, this implies
$\lim_{\lambda\rightarrow\infty}a(\lambda)\searrow0$ and so
\[
\lim_{\lambda\rightarrow\infty}\left(  \sup_{x\geq\overline{x}}V(x,\lambda
)-x\right)  =0.
\]
In addition, since for all $x\geq0$ such that $V_{x}(x,\lambda)$ exists,
$V_{x}(x,\lambda)\geq1$ we get that $h(x,\lambda):=V(x,\lambda)-x$ is
non-decreasing on $x$. As a consequence,
\[
V(x,\lambda)-x\leq V(\overline{x},\lambda)-\overline{x}=a(\lambda),
\]
so that we get the result. \hfill $\blacksquare$


\begin{thebibliography}{9}     
	
	\bibitem {AA06}Albrecher, H. and Asmussen, S. (2006). Ruin probabilities and aggregrate claims distributions for shot noise Cox processes. \textit{Scandinavian Actuarial Journal}, \textbf{2006}(2), 86-110.
	                                                                                           %
\bibitem {AAM24}Albrecher H., Azcue P. and Muler N. (2024). Optimal
dividend strategies for a catastrophe insurer. \textit{Frontiers of
	Mathematical Finance}, 3(2): 304-344.
	
\bibitem {ABT17}	Albrecher, H., Beirlant, J. and Teugels, J. (2017) Reinsurance: Actuarial and Statistical Aspects. John Wiley and Sons, Chichester.
	
	\bibitem {AT}Albrecher, H. and Thonhauser, S. (2009). Optimality results for
	dividend problems in insurance. \textit{RACSAM-Revista de la Real Academia de
		Ciencias Exactas, Fisicas y Naturales. Serie A. Matematicas} \textbf{103},
		
	\bibitem {AT12}	Albrecher, H. and Thonhauser, S. (2012). On optimal dividend strategies in insurance with a random time horizon. In: \textit{Stochastic Processes, Finance and Control: A Festschrift in Honor of Robert J Elliott}, pp. 157-179, World Scientific.
		
		\bibitem {AAU}Asmussen, S., Avram, F. and Usabel, M. (2002). Erlangian approximations for finite-horizon ruin probabilities. \textit{ASTIN Bulletin} 32(2), 267-281.
		
		\bibitem {avanzi}Avanzi, B. (2009). Strategies for dividend distribution: A
		review. \textit{North American Actuarial Journal} \textbf{13}, 2, 217-251.
		
		
		\bibitem {AM 2014}Azcue P. and Muler N. (2014). \textit{Stochastic
		Optimization in Insurance: a Dynamic Programming Approach. }Springer Briefs in
	Quantitative Finance. Springer.
	
	\bibitem {AM 2021}Azcue, P. and Muler N. (2021). A multidimensional problem of
	optimal dividends with irreversible switching: a convergent numerical scheme.
	\textit{Appl Math Optim} \textbf{83}, 1613--1649.
	
	\bibitem{BladtNielsen} Bladt, M. and Nielsen B.F. (2017) \textit{Matrix-exponential distributions in applied probability}. Vol. 81. New York: Springer.
	
\bibitem {Carr}Carr, P. (1998). Randomization and the American put. \textit{The Review of Financial Studies} 11(3), 597-626.

	\bibitem {Dassios Jang}Dassios, A., and Jang, J.W. (2003). Pricing of catastrophe
	reinsurance and derivatives using the Cox process with shot noise intensity.
	Finance Stochastics \textbf{7}, 73--95.
	
	\bibitem {defin}De Finetti, B. (1957). Su un'Impostazione Alternativa della
	Teoria Collettiva del Rischio. \textit{Transactions of the 15th Int. Congress
		of Actuaries} \textbf{2}, 433--443.
		
	\bibitem {Jiang}	Jiang, Z., and Pistorius, M. (2012). Optimal dividend distribution under Markov regime switching. Finance and Stochastics, \textbf{16}, 449-476.
		
\bibitem {LentonTipping} Lenton, T. M., Rockström, J., Gaffney, O., Rahmstorf, S., Richardson, K., Steffen, W., and Schellnhuber, H. J. (2019). Climate tipping points -- too risky to bet against. \textit{Nature} \textbf{575}(7784), 592-595.

\bibitem {filter} Leobacher, G., Szölgyenyi, M. and Thonhauser, S. (2014). Bayesian dividend optimization and finite time ruin probabilities. \textit{Stochastic Models} \textbf{30.2}, 216-249.


\bibitem {Lloyd}Lloyd, E., and Shepherd, T. (2020). Environmental catastrophes, climate change, and attribution. \textit{Annals of the New York Academy of Sciences} \textbf{1469}(1), 105-124.


\bibitem {Schneider}Schneider, R. (2023). Extreme events require new forms of financial collaboration to become more resilient. \textit{Environment Systems and Decisions} \textbf{43}(4), 544-554.

\bibitem{Seneviratne}Seneviratne, S., Zhang, X.,  Adnan, M.,  Badi, W., Dereczynski, C.,  Di Luca, A.,  Ghosh, S., Iskander, I., Kossin, J., Lewis, S., Otto, F., Pinto, I., Satoh, M., Vicente-Serrano, S., Wehner, M. and Zhou, B. (2021), {Weather and climate extreme events in a changing climate},
Climate Change 2021: The Physical Science Basis. Contribution of Working Group	I to the Sixth Assessment Report of the Intergovernmental Panel on Climate Change, 1513--1766, Cambridge University Press. 

\bibitem{Szölgyenyi} Szölgyenyi, M. (2015). Dividend maximization in a hidden Markov switching model. \textit{Statistics and Risk Modeling} \textbf{32}(3-4), 143-158.

\bibitem {WYW2}Wei, J., Yang, H., and Wang, R. (2010). Optimal reinsurance and dividend strategies under the Markov-modulated insurance risk model. \textit{Stochastic Analysis and Applications}, \textbf{28}(6), 1078-1105.

\bibitem {WYW}Wei, J., Yang, H., Wang, R. (2011). Optimal Threshold
Dividend Strategies under the Compound Poisson Model with Regime Switching.
In: Kohatsu-Higa, A., Privault, N., Sheu, SJ. (eds) \textit{Stochastic
	Analysis with Financial Applications. Progress in Probability}, vol 65.
Springer, Basel.
\end{thebibliography}
\end{document}